\documentclass[letterpaper, aps, prd, twocolumn, superscriptaddress, showpacs, nofootinbib]{revtex4}
\pdfoutput=1

\usepackage{graphicx}
\usepackage{bm}
\usepackage{dcolumn}
\usepackage{color}
\usepackage[sort&compress]{natbib}
\usepackage[latin9]{inputenc}
\usepackage{url}
\usepackage{subfigure}
\usepackage{float}
\usepackage{longtable}
\usepackage{amssymb}
\usepackage{amsmath}
\usepackage{amsfonts}

\begin{document}

\title{Inferring the core-collapse supernova explosion mechanism with three-dimensional gravitational-wave simulations }

\newcommand*{\glasgow}{University of Glasgow, Physics and Astronomy, Kelvin
Building, Glasgow, Lanarkshire, G12 8QQ, UK}
\affiliation{\glasgow}
\newcommand*{\ozgrav}{OzGrav, Swinburne University of Technology, Hawthorn, VIC 3122, Australia}
\affiliation{\ozgrav}
\newcommand*{\embryriddle}{Embry Riddle University, 3700 Willow Creek Road, Prescott Arizona, 86301, USA}
\affiliation{\embryriddle}

\author{Jade Powell}	    \affiliation{\glasgow} \affiliation{\ozgrav}
\author{Marek Szczepanczyk} \affiliation{\embryriddle}
\author{Ik Siong Heng}	    \affiliation{\glasgow}

\date{\today}

\begin{abstract} 
A detection of a core-collapse supernova signal with an Advanced LIGO and Virgo gravitational-wave detector network will allow us to measure astrophysical parameters of the source. In real advanced gravitational-wave detector data there are transient noise artifacts that may mimic a true gravitational-wave signal. In this paper, we outline a procedure implemented in the Supernova Model Evidence Extractor (SMEE) that determines if a core-collapse supernova signal candidate is a noise artefact, a rapidly-rotating core-collapse supernova signal, or a neutrino explosion mechanism core-collapse supernova signal. Further to this, we use the latest available three-dimensional gravitational-wave core-collapse supernova simulations, and we outline a new procedure for the rejection of background noise transients when only one detector is operational. We find the minimum SNR needed to detect all waveforms is reduced when using three-dimensional waveforms as signal models. 
\end{abstract}

\maketitle

\section{Introduction}
\label{sec:intro}

The first observing run of the Advanced Laser Interferometer Gravitational-Wave Observatory (aLIGO) \cite{aLIGO} began in September 2015 and made the first direct detections of gravitational waves \cite{GW150914, GW151226, o1cbcsearch}. The Advanced Virgo (AdV) gravitational-wave detector joined the advanced detector network during the second observing run in August 2017 \cite{AdVirgo}.  

Core-collapse supernovae (CCSNe) are expected to emit gravitational waves in the sensitivity range of advanced gravitational-wave detectors \cite{ott:04}. Although no CCSNe have currently been detected by gravitational-wave detectors, previous studies indicate that an advanced detector network may be sensitive to these sources out to the Large Magellanic Cloud (LMC) \cite{2016arXiv160501785A, gossan:16}. A CCSN would be an ideal multi-messenger source for aLIGO and AdV, as neutrino and electromagnetic counterparts to the signal would be expected. The gravitational waves are emitted from deep inside the core of CCSNe, which may allow astrophysical parameters, such as the equation of state (EOS), to be measured from the reconstruction of the gravitational-wave signal.

During the advanced detector observing runs, the Supernova Model Evidence Extractor (SMEE) is used as a parameter estimation follow-up tool for any potential gravitational-wave signal candidates that are identified by the gravitational-wave searches, or from alerts sent by electromagnetic and neutrino detectors \cite{powell:16b,logue:12}. The main purpose of SMEE is to identify the CCSN explosion mechanism. Zero age main sequence (ZAMS) stars, with masses $8\,\mathrm{M}_{\odot} < \mathrm{M} < 100\,\mathrm{M}_{\odot}$, form electron-degenerate cores. The star's nuclear burning stops when the core of the star is composed of iron nuclei, and then collapses when the star's core mass reaches the effective Chandrasekhar mass \cite{baron:90,bethe:90}. The collapse of the core will continue until the core reaches nuclear densities. The EOS stiffens above nuclear density, the inner core then rebounds, and a shock wave is launched outwards from the core. The shock then loses energy by nuclear dissociation and the emission of neutrinos from the optically thin regions. The shock then stalls and becomes an accretion shock, which must be revived within $\sim0.5-3\,$s, or the star will not explode, and will form a black hole as matter is accreted on to the proto-neutron star \cite{oconnor:11}. The explosion mechanism needed to revive the shock is currently not well understood, and may be determined from the reconstructed waveform of a CCSN signal detected with an advanced gravitational-wave detector network.

CCSNe simulations are difficult and computationally expensive, therefore only a small number of simulated CCSN gravitational-wave signals are available at this time. Common features have been identified in some of the available CCSN simulations. This allows us to identify signals, without robust knowledge of the phase, by associating different features with different explosion mechanisms. SMEE applies principal component analysis (PCA) via singular value decomposition to catalogues of CCSN waveforms. The principal components (PCs) can then be linearly combined to create signal models that represent each explosion mechanism. Bayesian model selection via nested sampling \cite{skilling:04, sivia:96}, can then be applied to determine the most likely explosion mechanism of the gravitational-wave signal. 

The first attempt to reconstruct a CCSN gravitational-wave signal without knowledge of the waveform was carried out by Summerscales \emph{et al.}\,\cite{summerscales:08}, and the first attempt to decompose a CCSN waveform catalogue into its main features was by Brady \emph{et al.} \cite{brady:04}, who used a Gram-Schmidt decomposition. Heng \cite{heng:09} was the first to apply PCA to CCSN waveforms, using waveforms from the Dimmelmeier \emph{et al.} \cite{dimmelmeier:08} waveform catalogue, and R{\"o}ver~\emph{et al.} \cite{roever:09} were the first to combine PCA with Bayesian data analysis techniques for CCSN waveform reconstruction. Similar techniques have been used to extract physical parameters of gravitational-wave signals from binary systems \cite{cannon:10, smith:13, 0264-9381-31-19-195010, PhysRevLett.115.121102, 0264-9381-33-8-085003}, and in characterizing noise sources in gravitational-wave detectors \cite{powell:15, powell:16}.

The first SMEE study (see Logue \emph{et al.}\,\cite{logue:12}) considered signals from neutrino-driven convection (the neutrino mechanism) \cite{murphy:09}, rapidly-rotating core-collapse (the magnetorotational mechanism) \cite{dimmelmeier:08}, and proto-neutron star pulsations (the acoustic mechanism) \cite{burrows:06,ott:04}. The signals were added, \textit{injected}, in Gaussian noise for a single aLIGO detector. A second study by Powell \emph{et al.} \cite{powell:16b} considered only signals from the neutrino and magnetorotational mechanisms using a three detector network. Real aLIGO and AdV noise taken during the initial detector science runs was used with the noise sensitivity altered to match the advanced detectors design sensitivity. Although the second SMEE study gave a more realistic estimate of how well the explosion mechanism can be determined, there are still some assumptions made in the analysis that do not match what we would expect from a real CCSN signal candidate. The second SMEE study assumes that it is already known that the signal belongs to one of the explosion mechanisms considered by SMEE. In real advanced gravitational-wave detector data there are transient noise artifacts, known as \textit{glitches}, that may mimic a true gravitational-wave signal. 

In this paper, we outline a new procedure that allows us to determine if a CCSN signal candidate is a glitch before the CCSN explosion mechanism is determined. Further to this, we update SMEE to use the latest available three-dimensional gravitational-wave core-collapse supernova simulations. Two-dimensional rapidly-rotating core-collapse waveforms are still a good approximation for three-dimensional rapidly-rotating waveforms, as non-axisymmetric instabilities occur after the signal bounce. However, three-dimensional neutrino mechanism waveforms exhibit different features to two-dimensional waveforms. Therefore, two-dimensional neutrino mechanism waveforms are not a good representation of what we expect to observe as astrophysical CCSNe are three-dimensional. 

This paper is structured as follows: In Section \ref{sec:waveforms}, we give a brief description of the CCSN waveforms used in this study. In Section \ref{sec:smee}, we describe the SMEE parameter estimation algorithm. In Section \ref{sec:no_PCs}, we determine the ideal number of PCs to use when creating the signal models for SMEE. In Section \ref{sec:noiseonly}, we show the result expected from SMEE when no signals are present in the data. In Section \ref{sec:minsnr}, we determine the minimum signal to noise ratio (SNR) needed for SMEE to detect CCSN waveforms. In Section \ref{sec:analysis}, we outline the analysis performed to test the improvements to the SMEE algorithm. In Section \ref{sec:results}, we provide the results for a single and multi detector analysis when signals are included in the model, and in Section \ref{sec:robust}, we provide the results when signals are not included in our model. A summary and discussion is provided in Section \ref{sec:discussion}.

\section{CCSN Waveforms}
\label{sec:waveforms}

\begin{figure*}[!ht]
\centering
\includegraphics[width=0.45\textwidth]{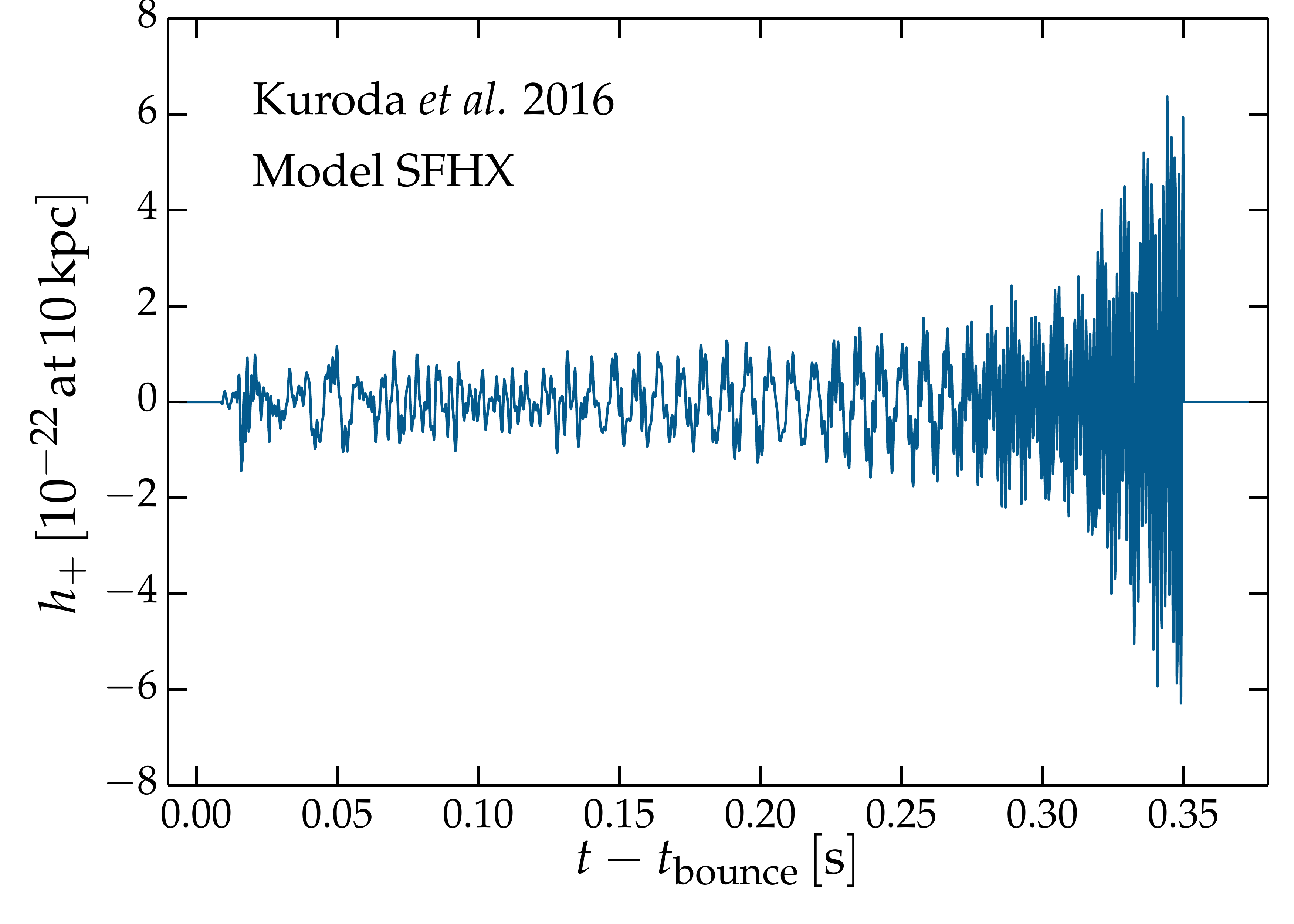}
\includegraphics[width=0.45\textwidth]{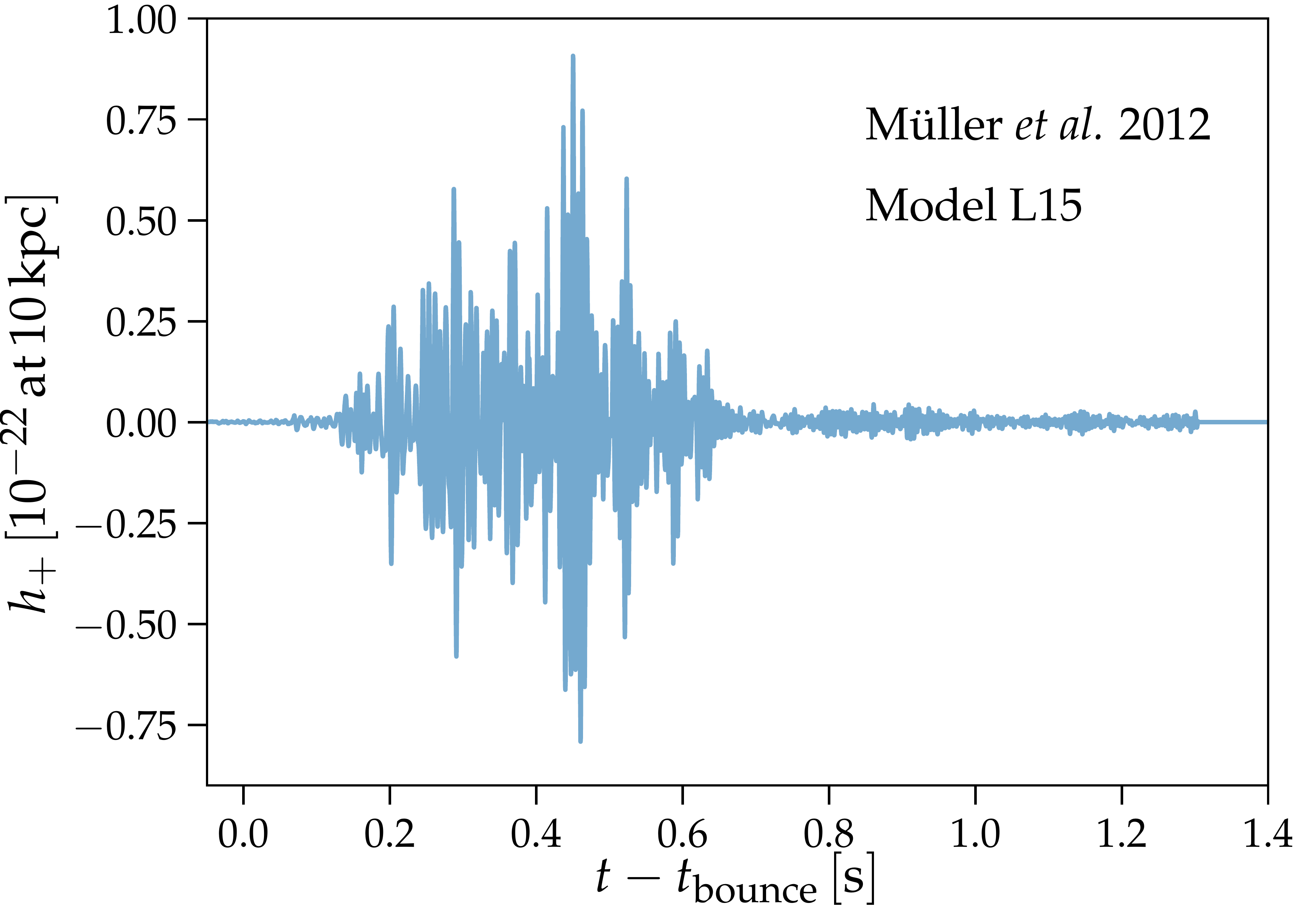}
\includegraphics[width=0.45\textwidth]{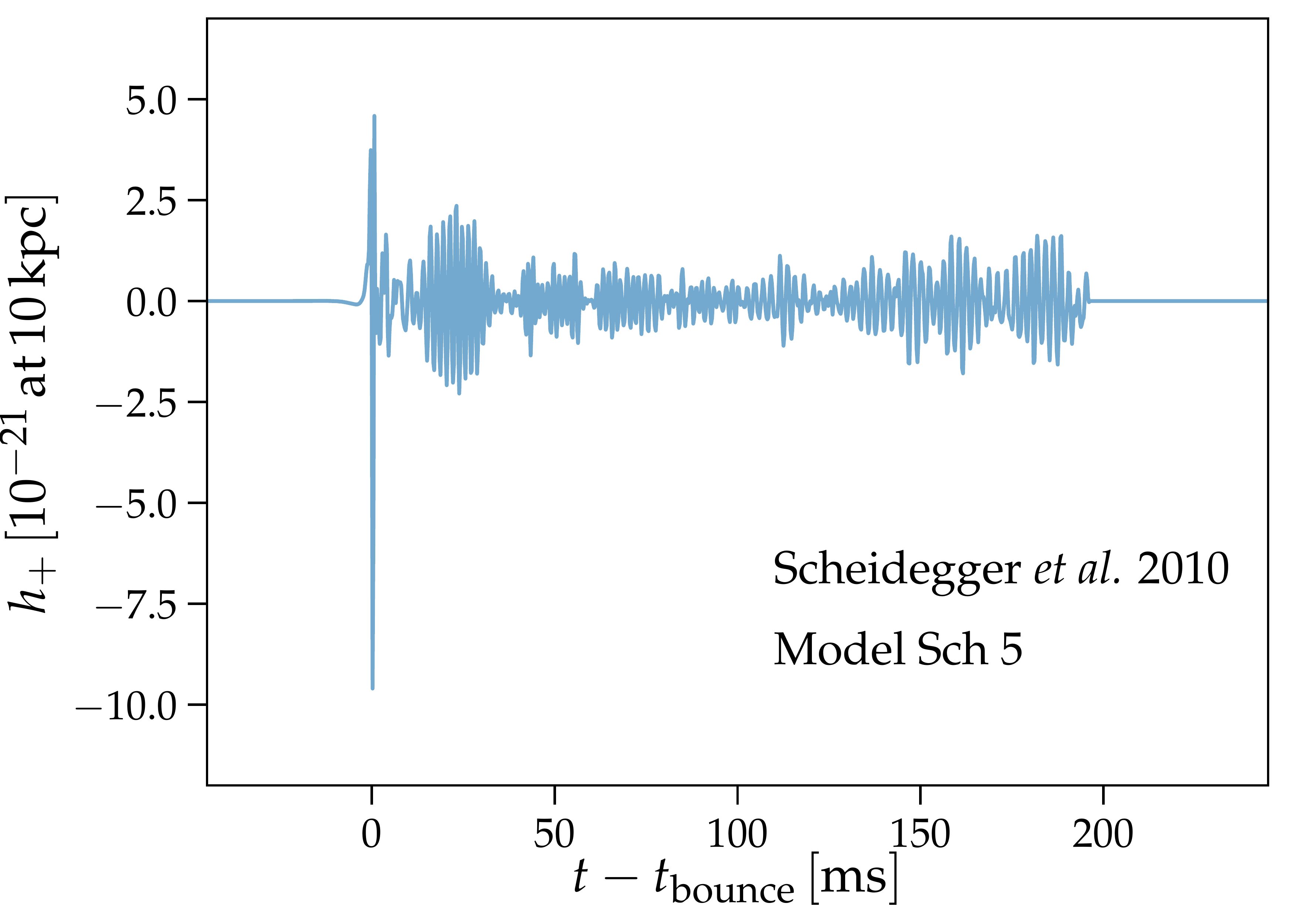}
\includegraphics[width=0.45\textwidth]{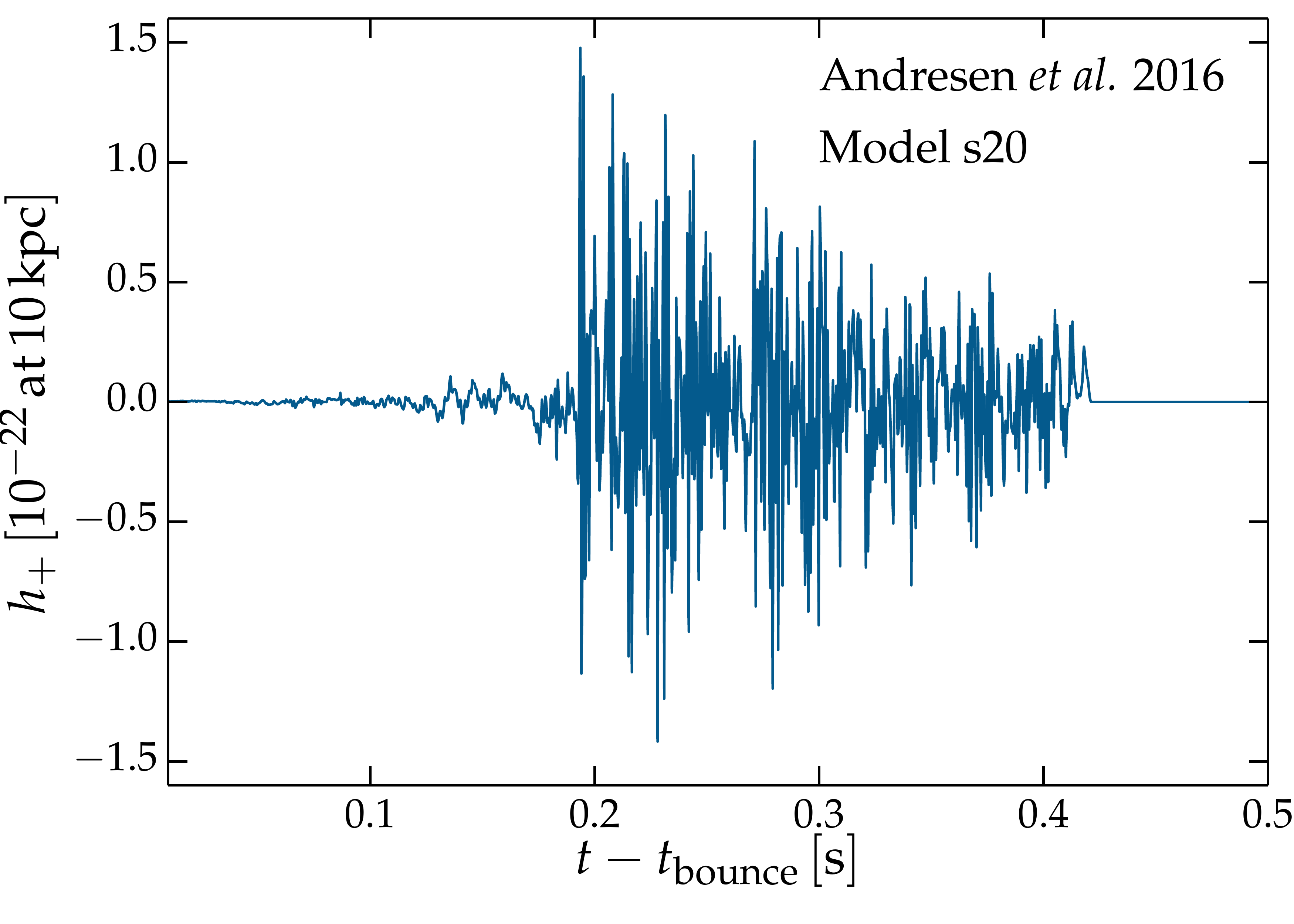}
\caption{Example gravitational-wave signals from CCSNe simulations. (Top left) Gravitational-waves from a neutrino mechanism $15\,\mathrm{M}_{\odot}$ progenitor star simulated by Kuroda \emph{et al} \cite{kuroda:16}. (Top right) A M\"uller \emph{et al.} \cite{mueller:e12} neutrino mechanism waveform with a $15\,\mathrm{M}_{\odot}$ progenitor star. (Bottom left) A representative magnetorotational waveform from Scheidegger \emph{et al.} \cite{scheidegger:10b} with a $15\,\mathrm{M}_{\odot}$ progenitor star. (Bottom right) A neutrino mechanism simulation with a $20\,\mathrm{M}_{\odot}$ progenitor star simulated by Andresen \emph{et al.} \cite{andresen:16}.}
\label{fig:waveforms}
\end{figure*}

In this paper, we consider two different possible explosion mechanisms for CCSNe. They are the magnetorotational explosion mechanism, where the explosion is powered by strong rotation and magnetic fields, and the neutrino explosion mechanism, where energy from neutrinos power the explosion. In this section, we give a brief description of the representative CCSNe waveforms from each mechanism that are used in this paper. 

\subsection{Neutrino mechanism}
\label{subsec:neutrino}

The neutrino mechanism is currently accepted as the most likely explosion mechanism for CCSNe.
Current reviews of the mechanism are given in \cite{janka:12a, burrows:13}. Neutrinos contain most of the energy, $\sim99\%$, released during the core-collapse \cite{bethe:90}. The neutrino mechanism involves some of the energy from the neutrinos being reabsorbed behind the initial shock to power the explosion. The gravitational-wave signals typically contain features produced by turbulence. This turbulence can be produced by convection and the standing accretion shock instability (SASI) \cite{blondin:03,foglizzo:07,foglizzo:06,scheck:08,kuroda:16}. The signals have a typical duration of $\sim0.3-2\,\mathrm{s}$, although some simulations are stopped early. The typical gravitational-wave energy of the explosions is $10^{-11} - 10^{-9}\,\mathrm{M}_{\odot}$, which leads to an amplitude of $\sim10^{-22}$ for a source at $10\,\mathrm{kpc}$.

Andresen \emph{et al.} \cite{andresen:16} produce four CCSN gravitational-wave signals for the first few hundred milliseconds after bounce. They are three-dimensional simulations, with multi-group neutrino transport, and ZAMS masses of $11.2\,\mathrm{M}_{\odot}$, $20\,\mathrm{M}_{\odot}$ and $27\,\mathrm{M}_{\odot}$. There are 3 failed explosions, one at each ZAMS mass, referred to as models s11, s20 and s27, and a successful explosion with a $20\,\mathrm{M}_{\odot}$ progenitor, referred to as model s20s. The gravitational-waves for the $11.2\,\mathrm{M}_{\odot}$ progenitor are convection-dominated, and the gravitational waves for the higher mass progenitors are SASI-dominated. They find that the SASI-dominated models are clearly distinguishable from the lower mass convection-dominated model by strong low-frequency emission between $100-200\,$Hz. They find that both the convection- and SASI-dominated models show gravitational-wave emission above $250\,$Hz, but their gravitational-wave amplitudes are much lower than previous two-dimensional simulations. The s20 model, scaled to a distance of $10\,$kpc, is shown in Figure \ref{fig:waveforms}. The gravitational-wave amplitude is larger for the exploding model. 

Kuroda \emph{et al.} \cite{kuroda:16} carry out fully relativistic three-dimensional simulations of a $15\,\mathrm{M}_{\odot}$ progenitor star with three different EOS. They are DD2 and TM1 \cite{2012ApJ...748...70H} and SFHx \cite{2013ApJ...774...17S}. We use two of these models which we refer to as SFHx and TM1. The simulations stop at around $\sim350\,$ms after bounce. Figure \ref{fig:waveforms} shows the waveform simulated with the SFHX EOS, scaled for a distance of $10\,$kpc. They find that the stiffness of the EOS creates significant changes in the SASI, and that the gravitational-wave frequency increases with time due to accretion on to the proto-neutron star. 

M\"uller~\emph{et al.}\,\cite{mueller:e12} performed three-dimensional simulations of three neutrino-driven CCSNe. They use grey neutrino transport and start the simulations after core bounce. There are three waveform models that we refer to as L15, W15 and N20. Their masses are $15\,\mathrm{M}_\odot$, $15\,\mathrm{M}_\odot$ and $20\,\mathrm{M}_\odot$, respectively. The L15 model is shown in Figure \ref{fig:waveforms}. Proto-neutron star convection occurs at late times in the signals.

\subsection{Magnetorotational mechanism}
\label{subsec:magnetorotational}

Rapidly-rotating CCSNe are highly energetic, and may be associated with high energy events, such as hypernovae and gamma ray bursts. Hypernovae are expected to be $\sim 1\%$ of all CCSNe \cite{2011MNRAS.412.1522S}. When core-collapse to a proto-neutron star occurs, it may result in spin-up of the stellar core by a factor of $\sim1000$ \cite{ott:06spin}. The rapidly-rotating pre-collapse core results in a millisecond period proto-neutron star, which if combined with a magnetar strength magnetic field could power a strong CCSN explosion. The main feature of rapidly-rotating gravitational-wave signals is a spike at core bounce. They are shorter in duration than the neutrino mechanism signals and generally an order of magnitude larger in amplitude.  

Scheidegger~\emph{et al.}\,\cite{scheidegger:10b} performed three-dimensional magnetohydrodynamical simulations of 25 gravitational-wave signals, using a leakage scheme for neutrino transport. They use a $15\,\mathrm{M}_{\odot}$ progenitor star, and the Lattimer-Swesty EOS \cite{lattimer:91}. An example Scheidegger waveform is shown in Figure \ref{fig:waveforms}. The waveforms contain only $\mathrm{h}_{+}$ around the spike at core bounce, and the $\mathrm{h}_{\times}$ polarization starts a few ms later. As Scheidegger \emph{et al.} use a variety of rotations in their simulations, from non-rotating to extremely rapidly-rotating, we discard the 10 simulations with the slowest rotation values leading to 15 rapidly-rotating waveforms available for use in SMEE. We refer to these waveforms as Sch1-15.  

\section{SMEE}
\label{sec:smee}

SMEE is designed as a parameter estimation follow-up analysis for possible detection candidates identified by gravitational-wave burst searches and external triggers, such as neutrinos, optical etc. SMEE is implemented in the LIGO data analysis software package \texttt{LALInference} \cite{veitch:15}, which is part of the LSC Algorithm Library (LAL) \cite{LAL}.

\subsection{Finding a signal}
\label{subsec:triggers}

To determine the explosion mechanism of a CCSN signal, SMEE will need to know the time at which the signal is expected to occur in the data. A detection of neutrinos from the source could give the time of the source with millisecond accuracy \cite{2006APh....26..155I, 2011NIMPA.656...11A}. It is possible that there could be a CCSN signal candidate that does not have any information from neutrinos. In this case it would not be possible to determine the time of the source with high levels of accuracy. 

If there is no neutrino signal then an on-source window can be calculated from electromagnetic observations of the supernovae. In the first gravitational-wave targeted search for CCSNe the on-source windows for the sources considered were 5.12 days, 1.26 days, 11.64 days and 1.52 days \cite{2016arXiv160501785A}. It would be too computationally expensive for SMEE to search for the signal over that length of time. In this case SMEE would follow-up triggers from the targeted supernova searches. It is also possible that a CCSN signal candidate could be identified by the all-sky un-modelled searches for gravitational-wave burst signals \cite{2016PhRvD..93d2004K, 0264-9381-25-11-114029, lynch:15}. 

In this paper, we assume the time of the signal is known due to a coincident neutrino detection. We also use search background triggers produced by Coherent WaveBurst (cWB) \cite{2016PhRvD..93d2004K, 0264-9381-25-11-114029}. The cWB algorithm whitens the data and converts to the time-frequency domain using the Wilson-Daubechies-Meyer wavelet transform \cite{1742-6596-363-1-012032}. Data from multiple detectors are then combined coherently to obtain a time-frequency power map. A signal is identified as a cluster of time-frequency data samples with power above some noise threshold. The signal waveforms in both detectors can then be reconstructed using a constrained likelihood method \cite{2005PhRvD..72l2002K}. More information about how cWB is used in the search for CCSNe can be found in \cite{2016arXiv160501785A}.

\subsection{Signal Models}
\label{subsec:models}

\begin{figure*}[ht]
\centering
\includegraphics[width=\textwidth]{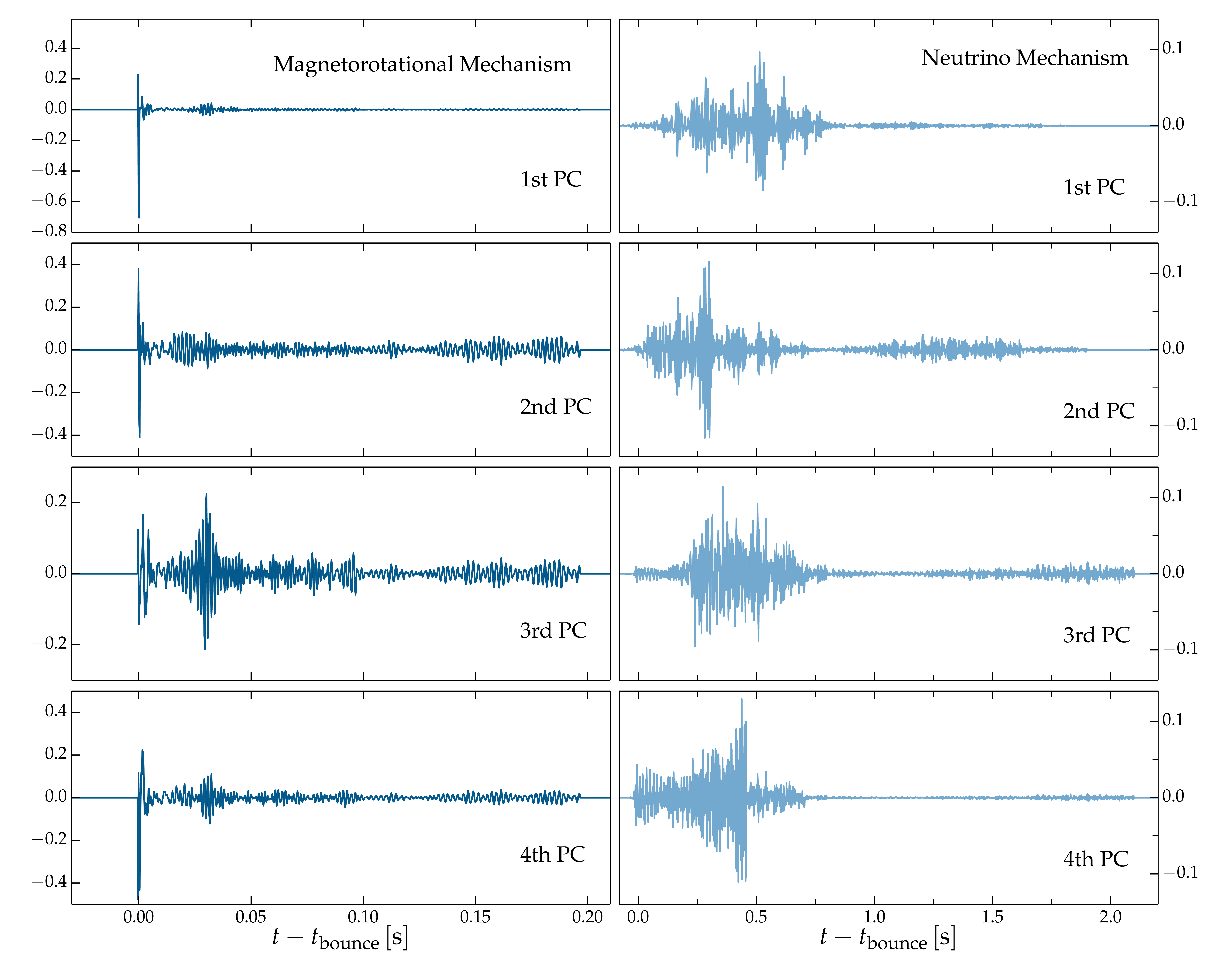}
\caption{The first four PCs used to represent each explosion mechanism. For the magnetorotational signal model, 15 waveforms from Scheidegger \emph{et al.}\,\cite{scheidegger:10b} are used. The main feature of the PCs is the spike at core bounce. The neutrino mechanism signal model is made from a mixture of waveforms simulated by Andresen \emph{et al.}\,\cite{andresen:16}, Kuroda \emph{et al.}\,\cite{kuroda:16} and M\"uller \emph{et al.}\,\cite{mueller:e12}. Turbulent features, such as convection and SASI, are the main features of the waveforms. Each PC is multiplied by an amplitude scale factor during the analysis.  } 
\label{fig:ch5_pcs}
\end{figure*}

SMEE is updated to use the latest available CCSN waveforms during the analysis of advanced gravitational-wave detector CCSN detection candidates. For the magnetorotational mechanism, the plus polarization of three-dimensional waveforms from  Scheidegger \emph{et al.} \cite{scheidegger:10b} are used. The waveforms from Scheidegger \emph{et al.} are $\sim100$\,ms longer in duration than the two-dimensional waveforms used in the previous SMEE papers \cite{logue:12,powell:16b}, allowing SMEE to reconstruct a longer part of the signal.

For the neutrino mechanism signal model, the plus polarization is used from a combination of three-dimensional waveforms simulated by M\"uller \emph{et al.} \cite{mueller:e12}, Andresen \emph{et al.} \cite{andresen:16}, and Kuroda \emph{et al.} \cite{kuroda:16}. There are three waveforms from M\"uller \emph{et al.} \cite{mueller:e12}, four waveforms from Andresen \emph{et al.} \cite{andresen:16}, and two waveforms from Kuroda \emph{et al.} \cite{kuroda:16}. This gives a total of 9 three-dimensional waveforms used to make the neutrino mechanism signal models. Any new waveforms that become available before the gravitational-wave detectors reach their design sensitivity will be added to the relevant models. Increasing the number of available waveforms will increase the accuracy of SMEE.

Singular value decomposition is applied to a matrix containing the time-series waveforms to identify the most important features of the different explosion mechanisms. As for the two-dimensional PCs used in \cite{powell:16b}, created from waveforms simulated by Dimmelmeier \emph{et al.} \cite{dimmelmeier:08}, the Scheidegger waveforms are aligned at the spike at core bounce before PCA is applied. The three-dimensional neutrino mechanism waveforms are aligned at the start of the gravitational-wave emission, with $t=0$ as shown in Figure 1. This is because it is difficult to identify common features in the time series that are a good place to align the waveforms. This can be changed in the future when SMEE is updated to use spectrograms. In a spectrogram it is easier to see common features, such as g-modes, that can be used to align waveforms. Both sets of waveforms are then zero padded to make them the same length. We search over three seconds of data, as this is large enough to account for differences in the definition of t=0 between the different models, the neutrino triggers and gravitational wave search triggers. 

PCA can be used to transform a collection of waveforms into a set of orthogonal basis vectors called principal components (PCs). The first few PCs represent the main features of a set of waveforms, therefore allowing for a dimensional reduction of the data set. By applying SVD to the original data matrix $D$, where each column corresponds to a supernova waveform, the data can be factored such that
\begin{align}
D &= U\,\Sigma\,V^{\mathrm{T}}\,,
\end{align}
where $U$ and $V$ are matrices of which the columns are comprised of the eigenvectors of $DD^{\mathrm{T}}$ and $D^{\mathrm{T}}D$, respectively. $\Sigma$ is a diagonal matrix with elements that correspond to the square root of the eigenvalues. The orthonormal eigenvectors in $U$ are the PCs. As the PCs are ranked by their corresponding eigenvalues in $\Sigma$, the main features of the data set are contained in just the first few PCs. Each waveform, $h_{i}$, in the data set can be reconstructed using a linear combination of the PCs, multiplied by their corresponding PC coefficients. A more detailed description of how SMEE works is given in the previous SMEE studies \cite{logue:12,powell:16b}.

The first four PCs for each mechanism are shown in Figure \ref{fig:ch5_pcs}. The main feature of the magnetorotational PCs is the spike at core bounce. Most of the gravitational-wave emission in the neutrino mechanism PCs is contained in the first $1\,$s. 

\subsection{Model selection}
\label{subsec:modelselec}

Bayesian model selection is used to calculate Bayes
factors that allow us to distinguish between two competing models. The Bayes factor, 
$B_{S,N}$, is given by the ratio of the evidences,
\begin{align}
B_{S,N} = \frac{p(D|M_{S})}{p(D|M_{N})}\,,
\end{align}
where $M_{S}$ and $M_{N}$ are the signal and noise models, respectively. The 
evidence is given by the integral of the likelihood multiplied by the prior 
across all parameter values. 
For a large number of parameters the evidence integral can become difficult.
This problem is solved using a technique known as nested sampling. A more detailed
description of nested sampling can be found in \cite{sivia:96,veitch:15,2010PhRvD..81f2003V}.
The log signal vs. noise Bayes factors for the individual models can be subtracted to obtain a new log Bayes factor ($\mathrm{logB}_{MagRot-Neu}$) that determines the explosion mechanism. If $\mathrm{logB}_{MagRot-Neu}$ is positive then the signal is a magnetorotational mechanism signal, and if $\mathrm{logB}_{MagRot-Neu}$ is negative then the signal is a neutrino mechanism signal. 

\subsection{Rejecting noise transients}
\label{subsec:reject}

A test that can be performed to determine if the signal is a real astrophysical signal or a glitch is the Bayesian coherence test \cite{2010PhRvD..81f2003V}. To perform the coherence test, we calculate a coherent vs. incoherent signal or noise Bayes factor $B_{C,IN}$. First, the evidence must be calculated coherently, as in the previous SMEE analysis. The coherent evidence $Z_{C}$, with combined data d from all detectors, which contains a coherent signal $S_{C}$ with parameters $\theta$, is given by,
\begin{equation}
Z_{C} = \int_{\theta} p(\theta|S_{C}) p(d|\theta,S_{C}) d\theta\,.
\end{equation} 
For an incoherent signal, each of the N detectors contains a signal which can be described by different parameters
for each detector. The incoherent evidence $Z_{I}$ is then given by,
\begin{equation}
Z_{I} = \prod_{j=1}^{N} \int_{\theta} p(\theta_{j}|S_{j}) p(d_{j}|\theta_{j},S_{j}) d_{j}\theta\,.
\end{equation}
The coherent vs. incoherent Bayes factor $B_{C,I}$ can then be calculated as,
\begin{equation}
B_{C,I} = Z_{C} / Z_{I}\,.
\end{equation} 
This can be extended to include a noise model. The incoherent or noise evidence $Z_{IN}$ is given by,
\begin{equation}
Z_{IN} = Z_{S}^{1}Z_{S}^{2} + Z_{S}^{1}Z_{N}^{2} + Z_{N}^{1}Z_{S}^{2} + Z_{N}^{1}Z_{N}^{2} \,,
\end{equation}
where $Z_{S}^{1}$ is the signal evidence for detector 1, $Z_{N}^{1}$ is the noise evidence for detector 1, $Z_{S}^{2}$ is the signal evidence for detector 2, and $Z_{N}^{2}$ is the noise evidence for detector 2. 
The coherent vs. incoherent or noise Bayes factor $B_{C,IN}$ is then given by, 
\begin{equation}
B_{C,IN} = Z_{C} / Z_{IN}\,.
\end{equation}

It is only possible to perform the coherence test when at least two gravitational-wave detectors are operational. When only a single detector is operational, SMEE can distinguish between glitches and CCSN signals by creating signal models for typical glitches. This is achieved by applying the same PCA technique applied to the CCSN waveforms to the time series of glitches found in the detector noise. Bayes factors can then be calculated to determine if the signal candidate is a CCSN or a glitch \cite{powell:15, powell:16}. This is a new approach that has not previously been used to reject background signals in single gravitational-wave detectors. 

\section{Determining the ideal number of PCs} 
\label{sec:no_PCs}

\begin{figure*}[!t]
\centering
\subfigure[]{\label{fig:pick_a}\includegraphics[width=0.45\textwidth]{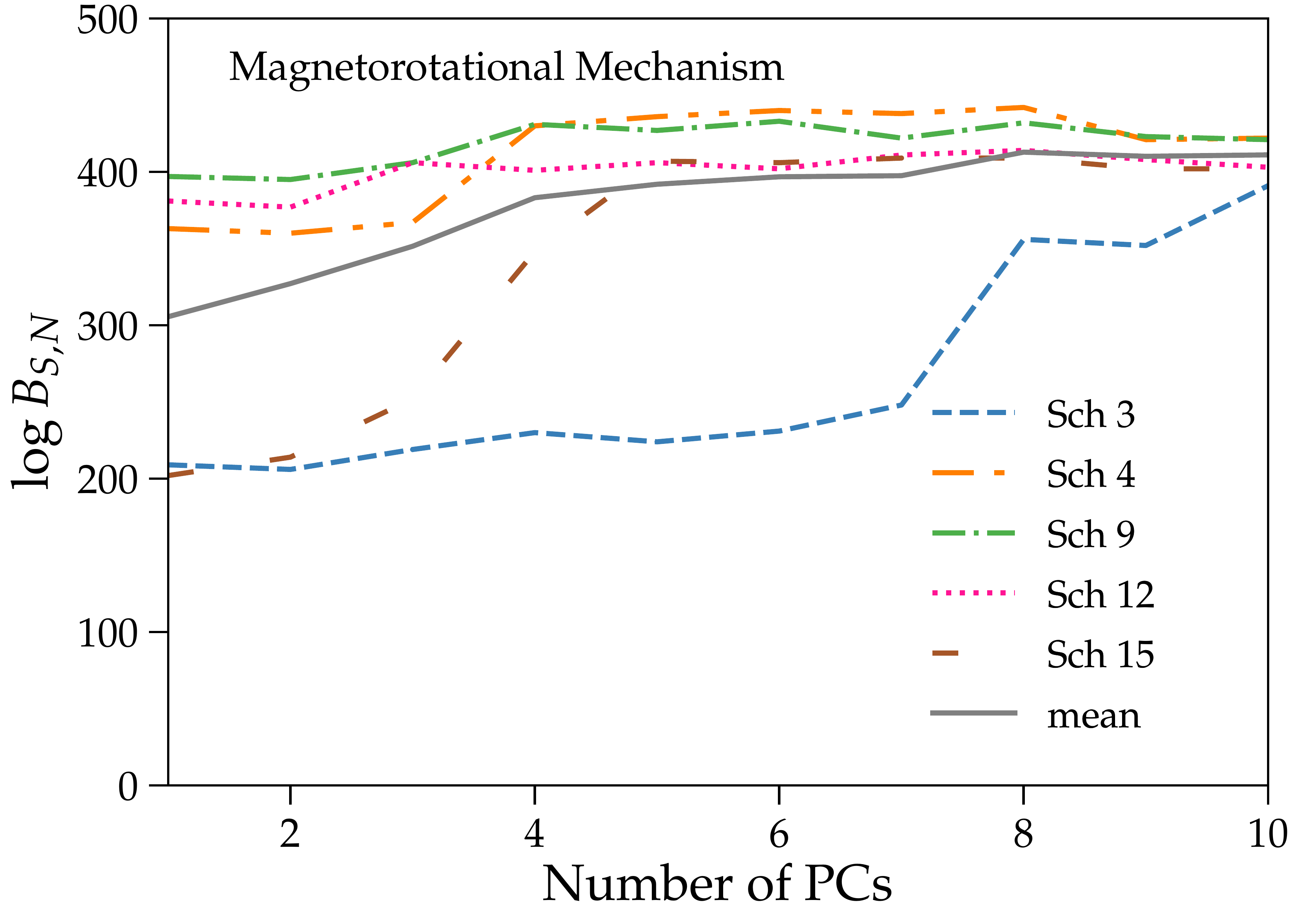}}
\subfigure[]{\label{fig:pick_b}\includegraphics[width=0.45\textwidth]{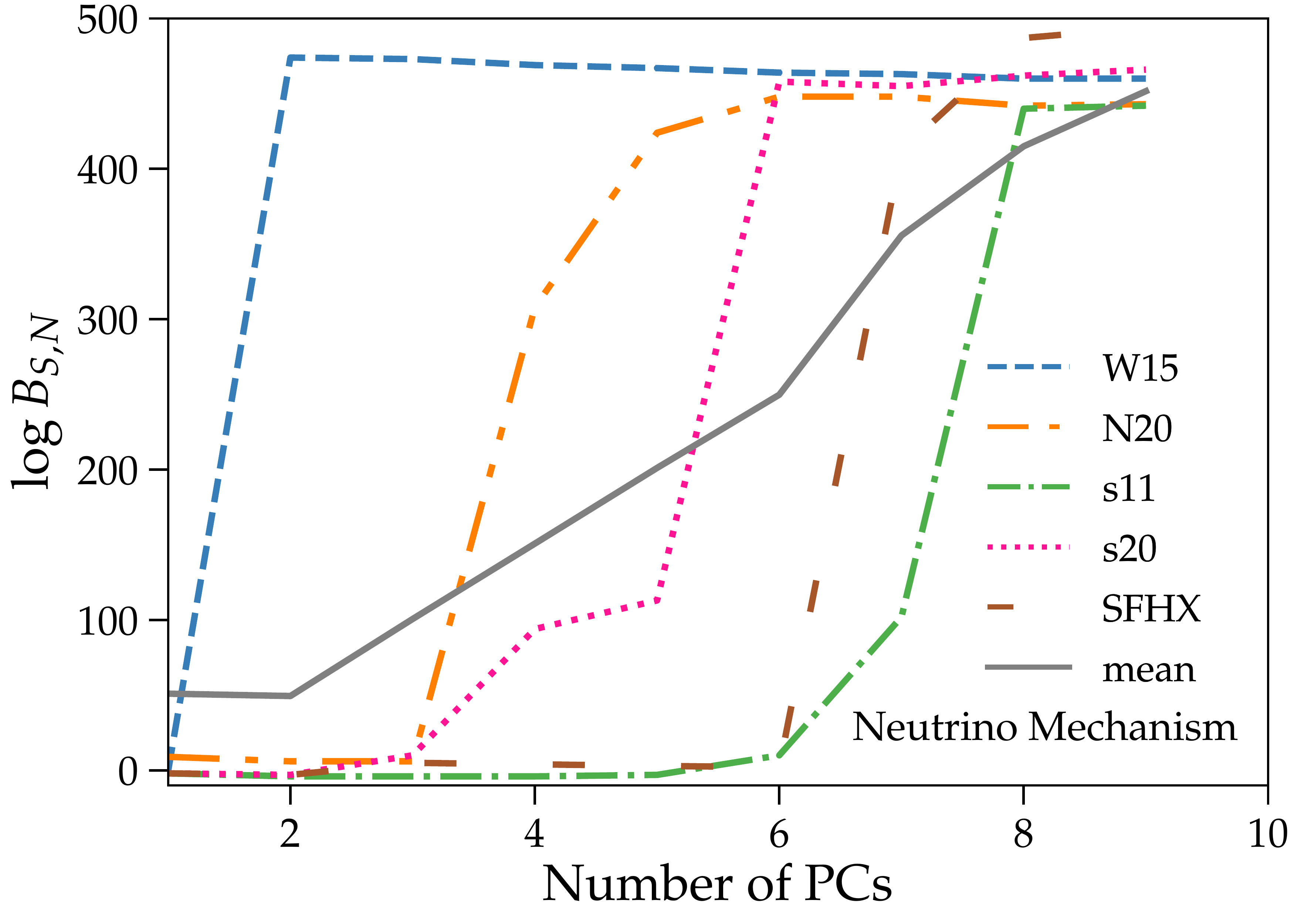}}
\subfigure[]{\label{fig:pick_c}\includegraphics[width=0.45\textwidth]{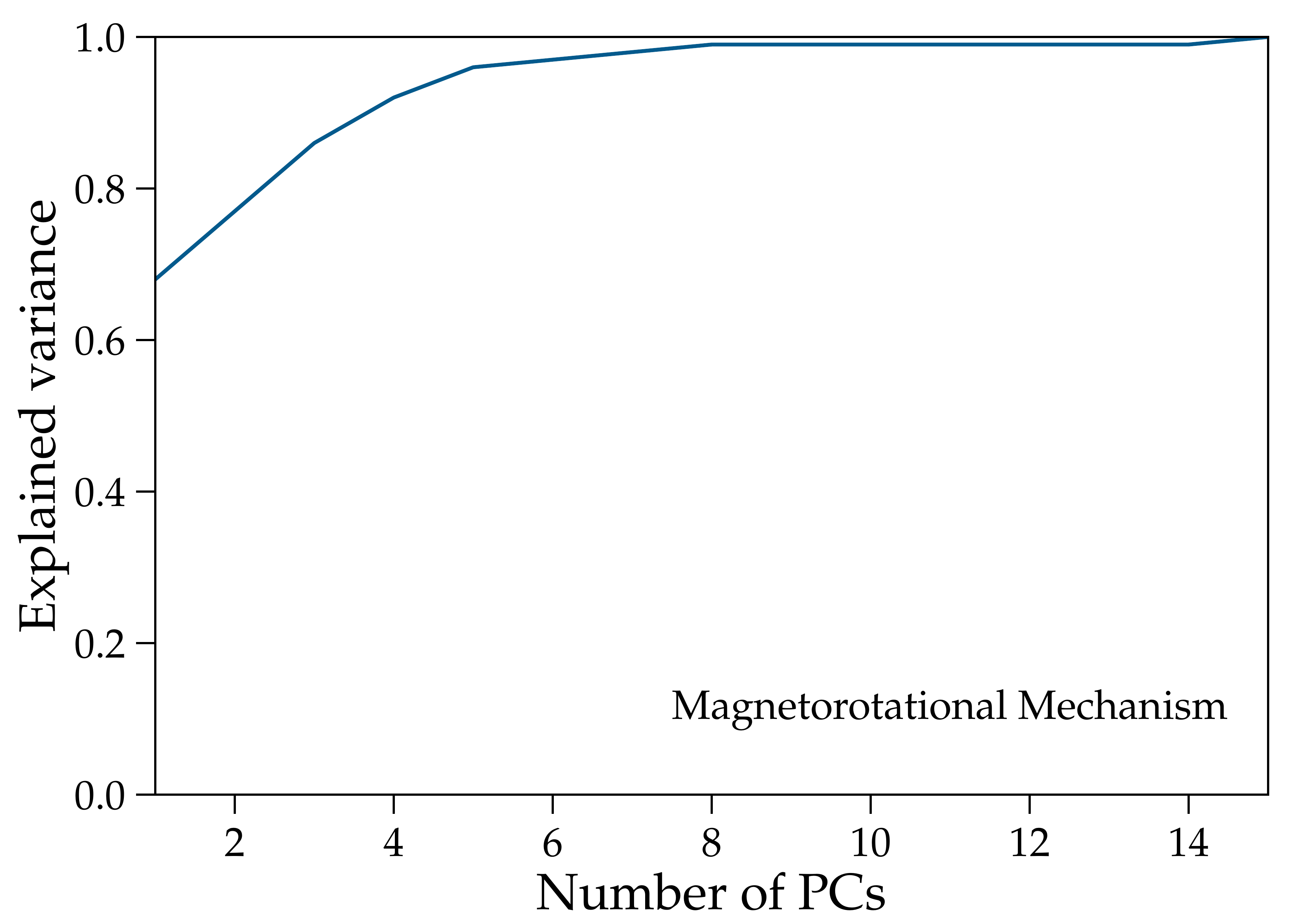}}
\subfigure[]{\label{fig:pick_d}\includegraphics[width=0.45\textwidth]{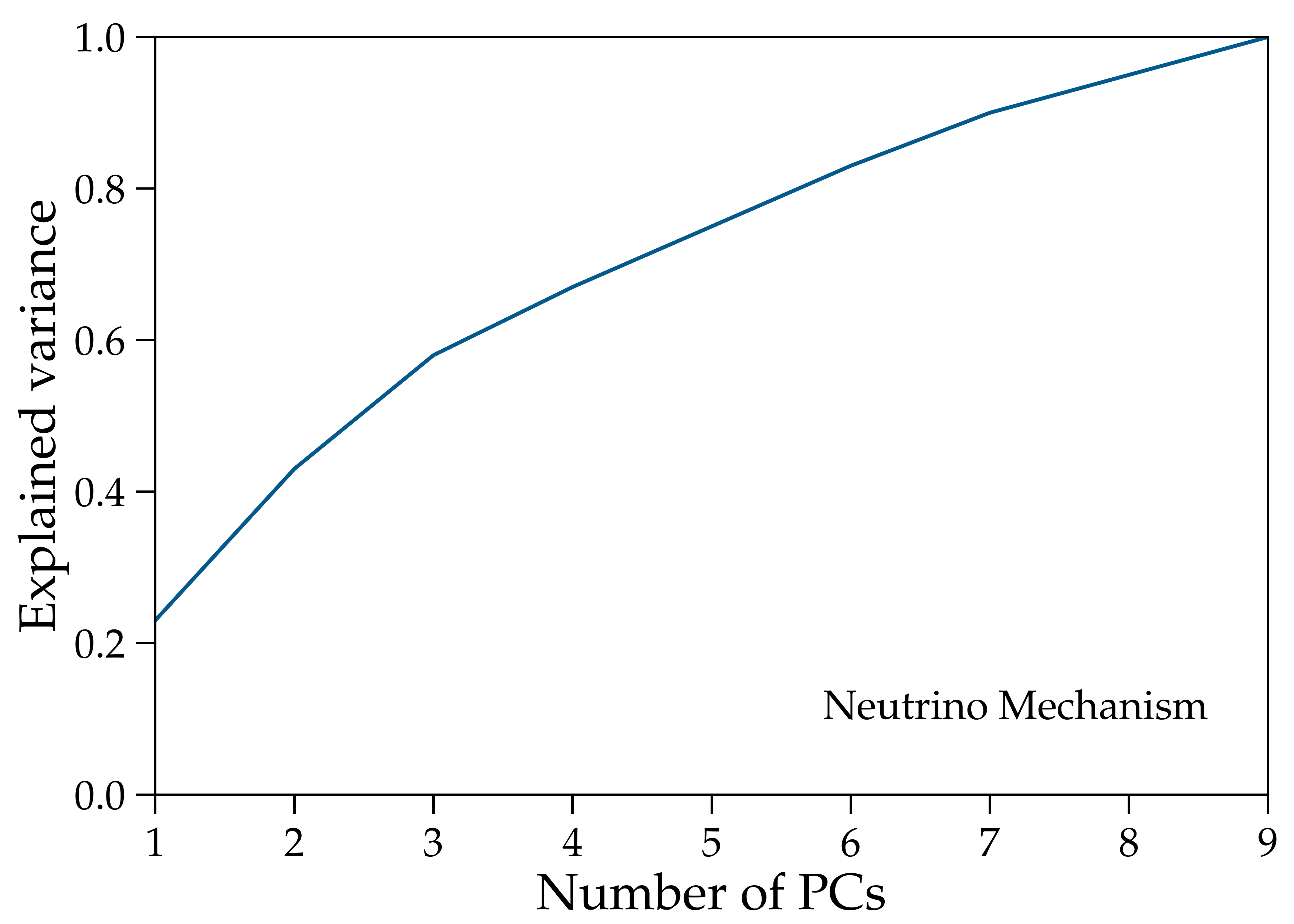}}
\caption{The increase in log Bayes factors and cumulative variance as the number of PCs is increased. \subref{fig:pick_a} The signal vs. noise log Bayes factors for an increasing number of PCs, for five representative magnetorotational waveforms. An ideal number of PCs is reached when the log Bayes factors are no longer sharply increasing. \subref{fig:pick_b} The same result for five representative three-dimensional neutrino mechanism signals. All signals are injected with an SNR of 17. \subref{fig:pick_c} The variance curve for the magnetorotational waveforms, and \subref{fig:pick_d} is the variance curve for the neutrino mechanism waveforms. We choose 5 PCs for the magnetorotational mechanism as this corresponds to the 'knee' of the variance curve. We choose 8 PCs for the neutrino mechanism as this value gives a high Bayes factor for all neutrino mechanism waveforms. }
\label{fig:ch5_pickpcs}
\end{figure*}

To determine the ideal number of PCs, the same method used in \cite{powell:16b} is applied to the new PCs used in this study. Figure \ref{fig:ch5_pickpcs}\subref{fig:pick_a} shows how the log signal vs. noise Bayes factors increase as the number of PCs is increased for the magnetorotational model. Five representative waveforms are selected from the Scheidegger \emph{et al.} waveform catalogue that span the full parameter range of the catalogue. Each of the waveforms are injected with a network SNR of 17. We use 24 hours of detector noise from the Initial LIGO S5 science run \cite{2009RPPh...72g6901A} and the Virgo VSR2 run, which has been recoloured to match the expected design sensitivity of the detectors \cite{vallisneri:15, gossan:16}. The signal vs. noise log Bayes factors increase slowly as more PCs are used in the signal model. There is no clear knee in the curves, which makes it difficult to determine from the figure what the ideal number of PCs is. The explained variance curve for the Scheidegger \emph{et al.} waveforms is shown in Figure \ref{fig:ch5_pickpcs}\subref{fig:pick_c}. The explained variance is the normalized cumulative sum of the eigenvalues calculated whilst making the PCs. It tells us how much of the total variance in the entire data set is encompassed in each PC. We use 5\,PCs for the magnetorotational explosion mechanism, as the variance encompassed in each PC increases more slowly after this number.      

Figure \ref{fig:ch5_pickpcs}\subref{fig:pick_b} shows how the signal vs. noise log Bayes factors increase as the number of PCs is increased for the neutrino mechanism signal model. As for the Scheidegger waveforms, five representative waveforms that span the parameter space are injected with an SNR of 17 in the recoloured detector noise. The injected waveforms are W15, N20, s11, s20 and SFHX.  The number of PCs needed for a good reconstruction of the waveforms has a lot of variety between the different waveforms. The W15 model requires 2 PCs, and the s11 model requires 8 PCs to achieve the optimal log Bayes factor. This means that the different three-dimensional neutrino mechanism waveforms do not share many common features in their time-series morphologies. The variance curve for the neutrino mechanism PCs is shown in Figure \ref{fig:ch5_pickpcs}\subref{fig:pick_d}. There is no clear change in the variance curve that could indicate the ideal number of PCs. So that all waveforms will be well represented by the neutrino mechanism signal model in SMEE, 8\,PCs are used for the three-dimensional neutrino mechanism signal model. 

\section{Noise only}
\label{sec:noiseonly}

As the signal models in SMEE are now created using three-dimensional CCSN waveforms, the result expected in the absence of a signal has changed since the previous result presented in \cite{powell:16b}. Therefore, in this section we recalculate the result expected from SMEE when no signals are present in the data. To achieve this, we run SMEE using both signal models on 1000 instances of Gaussian aLIGO and AdV design sensitivity noise, and 1000 instances of recoloured noise. 

\begin{figure}[h!]
\centering
\includegraphics[width=\columnwidth]{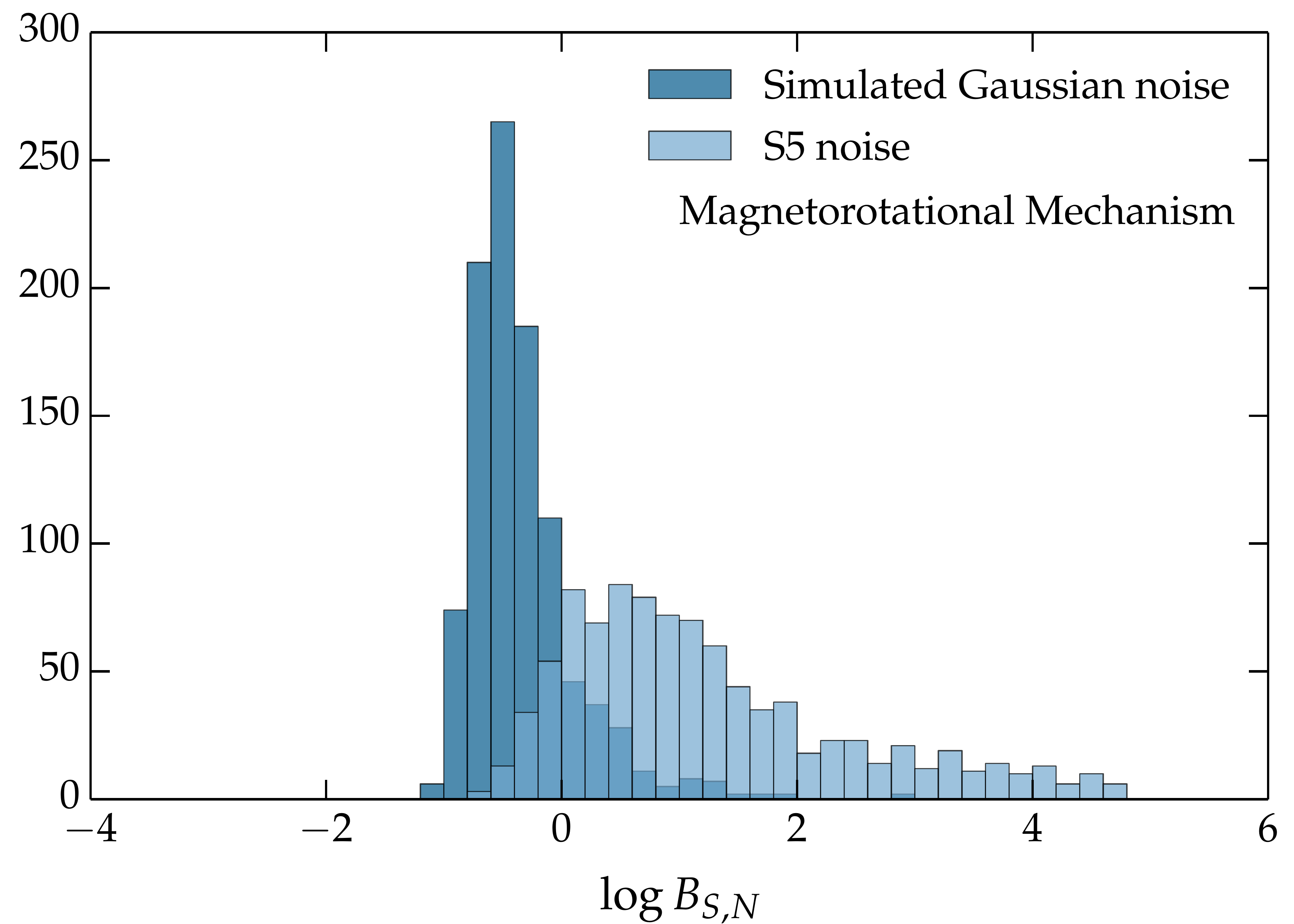}
\includegraphics[width=\columnwidth]{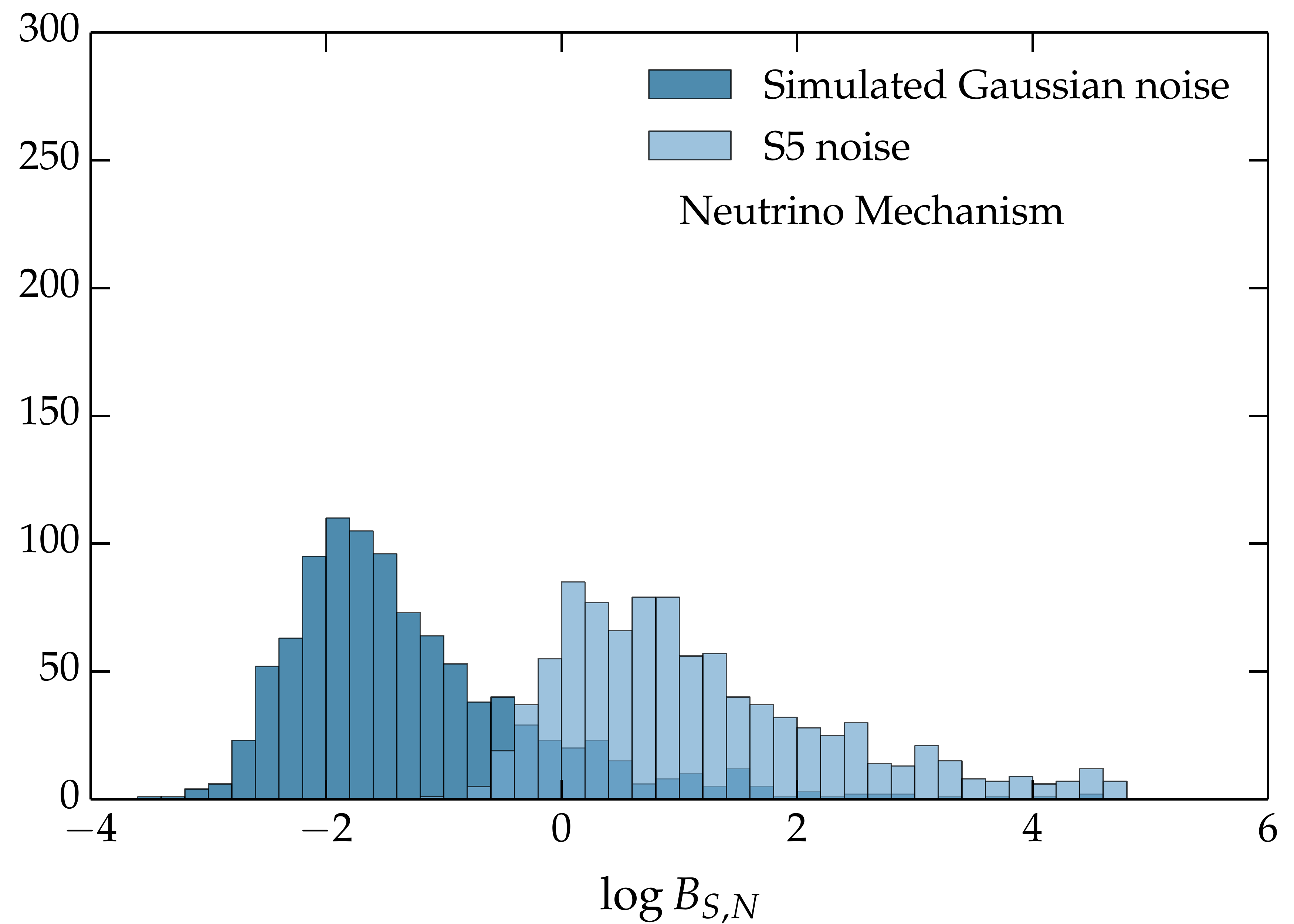}
\caption{The distributions of Bayes factors for 1000 instances of simulated aLIGO and AdVirgo design sensitivity noise and real recoloured noise. In Gaussian noise, when no signal is present we expect SMEE to produce a signal vs. noise log Bayes factor of $\sim-1$ or $\sim-2$ using the magnetorotational and neutrino signal models, respectively. In real noise, when no signal is present we expect SMEE to produce a signal vs. noise log Bayes factor of $\sim0$ using the magnetorotational or neutrino signal models. } 
\label{fig:ch5_noise_only}
\end{figure}

The results are shown in Figure \ref{fig:ch5_noise_only}. In Gaussian noise, when no signal is present we expect SMEE to produce a signal vs. noise log Bayes factor of $\sim-1$ and $\sim-2$ using the magnetorotational and neutrino signal models, respectively. In the recoloured noise, when no signal is present we expect SMEE to produce a signal vs. noise log Bayes factor of $\sim0$ using the magnetorotational or neutrino signal model. The log Bayes factors are higher in real detector noise due to the non-Gaussian, non-stationary noise features. Due to the spread in the Bayes factors, when no signal is present we only consider a signal to be detected if $\mathrm{logB}_{S,N} > 10$.

\section{Minimum SNR}
\label{sec:minsnr}

\begin{figure*}[!t]
\centering
\includegraphics[width=0.45\textwidth]{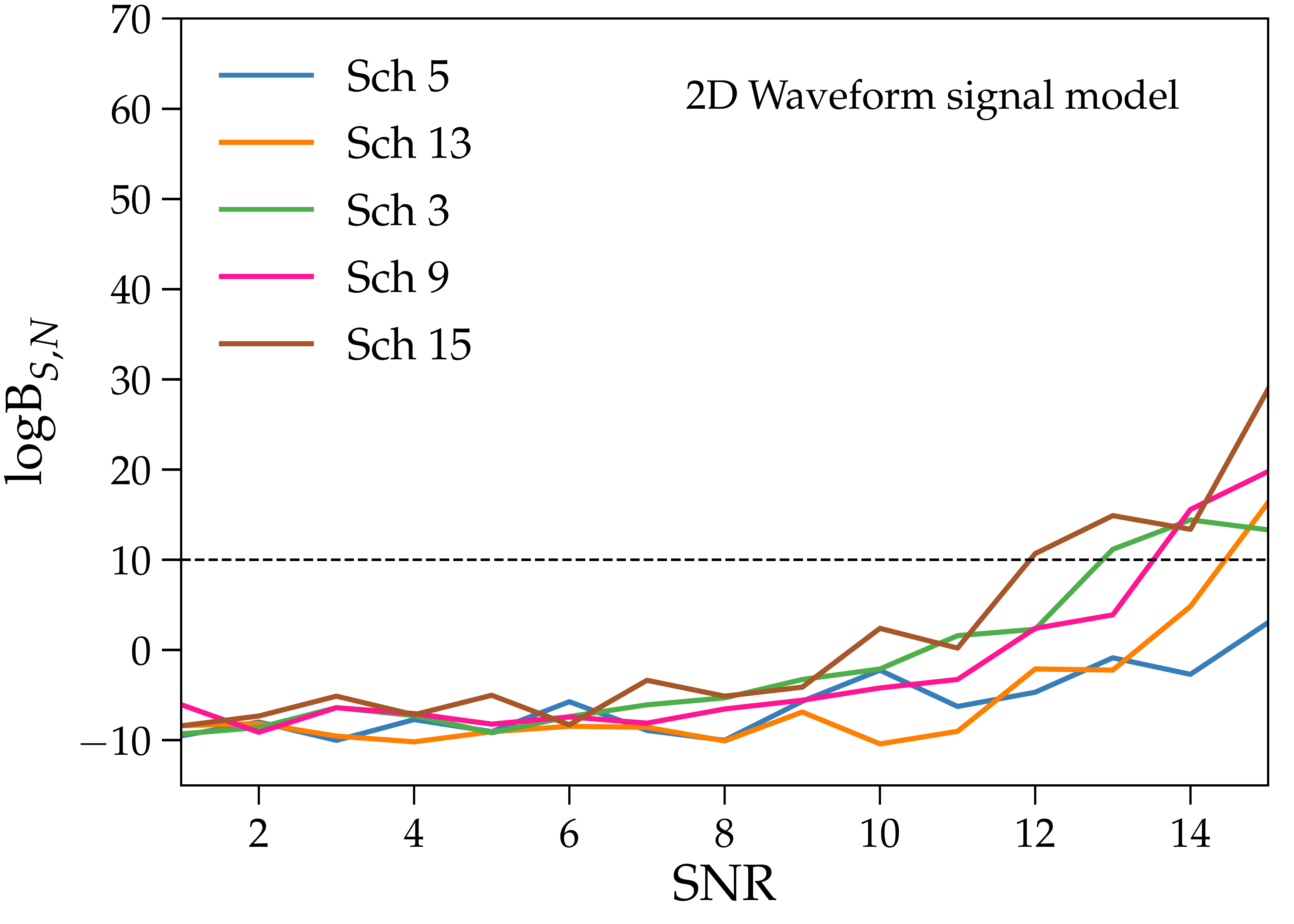}
\includegraphics[width=0.45\textwidth]{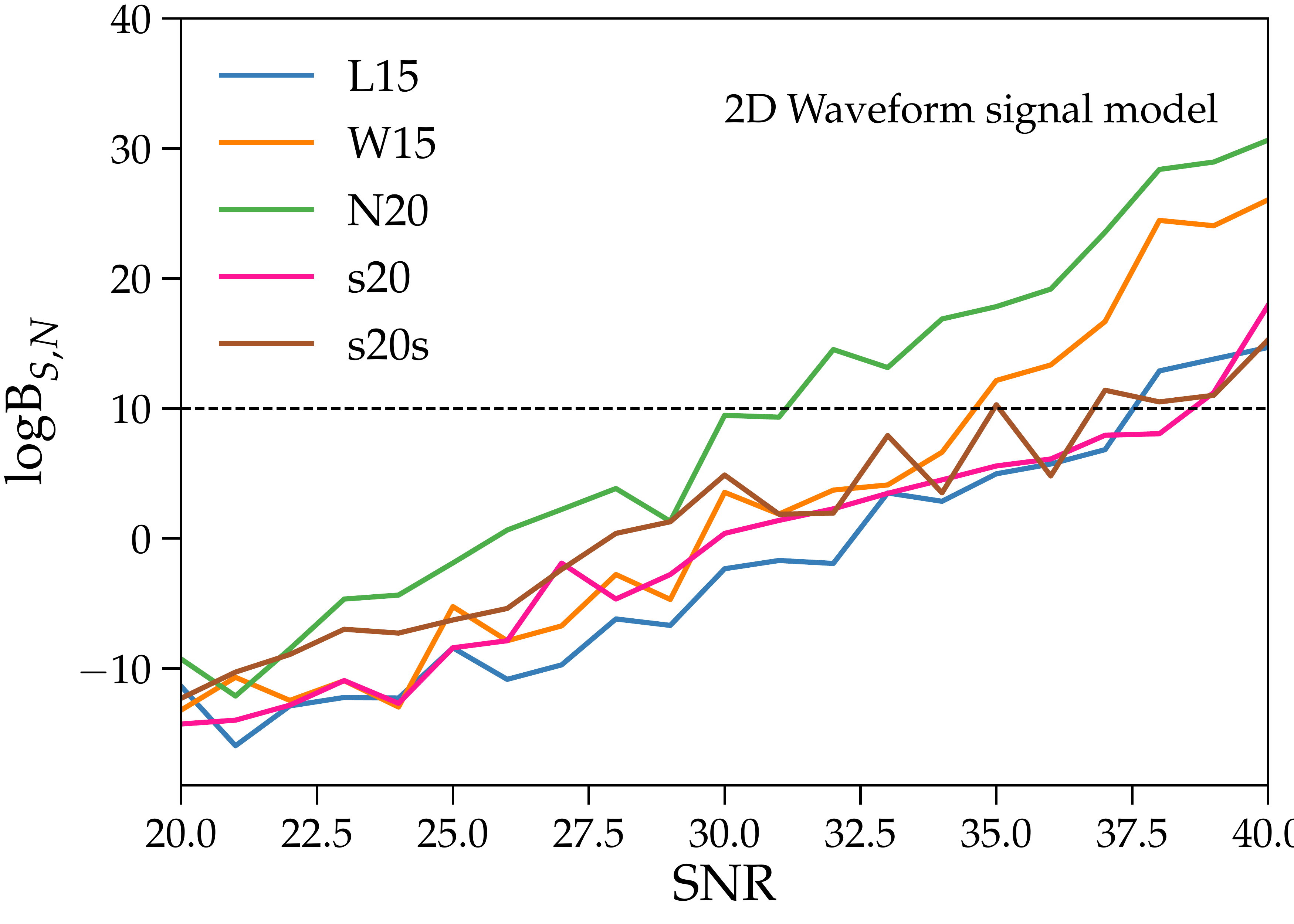}
\includegraphics[width=0.45\textwidth]{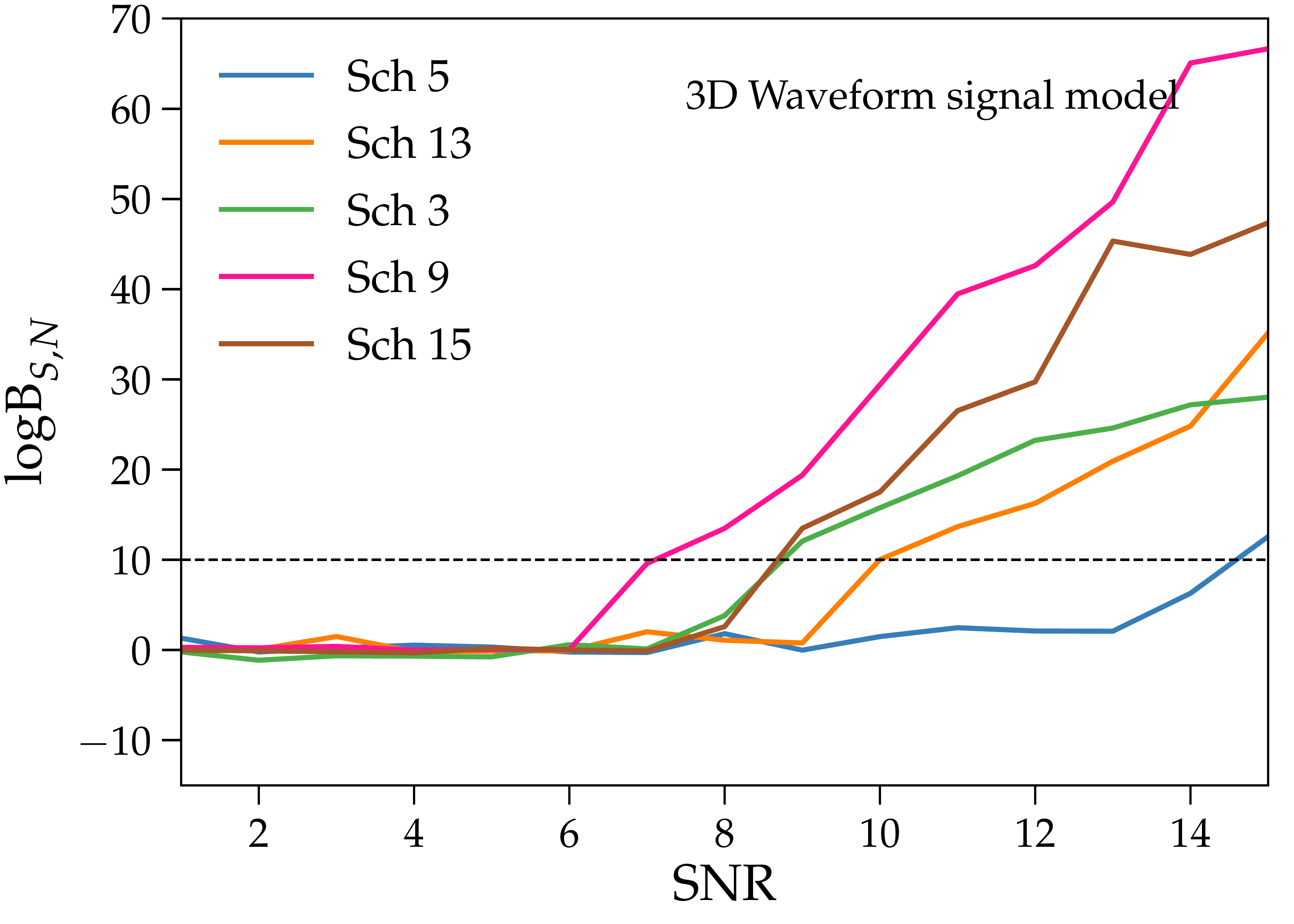}
\includegraphics[width=0.45\textwidth]{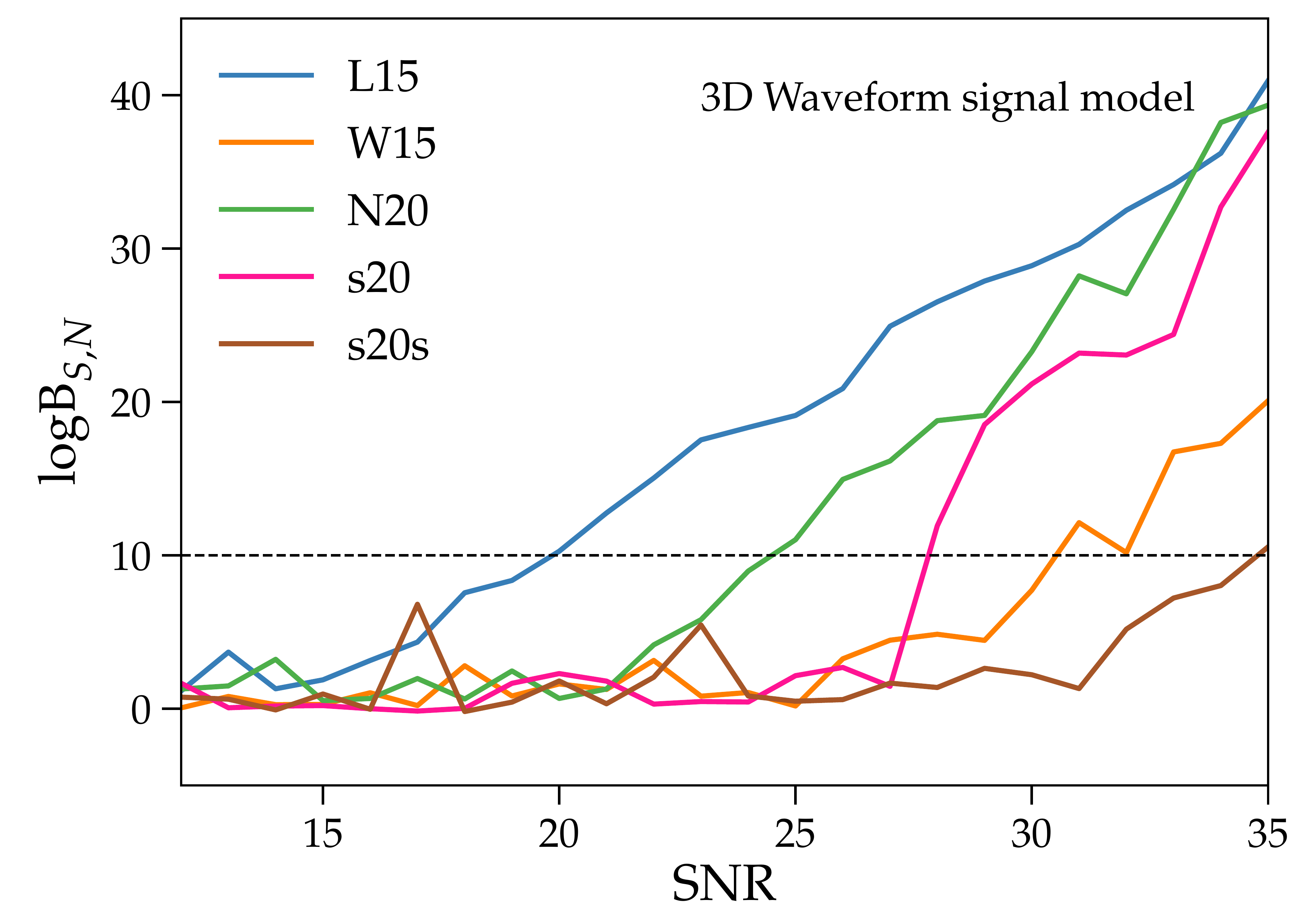}
\caption{The minimum network SNR needed to detect different CCSN signals. Five representative waveforms from each explosion mechanism are injected at increasing network SNR values and a Bayes factor is determined. In the top two figures we use the signal models created from the two-dimensional waveforms in \cite{powell:16b}. In the bottom two figures we use the new signal models created from three-dimensional waveforms. The dashed line shows the threshold for a detection. We find the minimum SNR needed to detect all models is improved by using the three-dimensional waveforms. }  
\label{fig:ch5_min_SNR}
\end{figure*} 

Understanding the minimum network SNR needed by SMEE is important, as the explosion mechanism cannot be determined for a signal that is not detected. The minimum network SNR needed for SMEE to detect CCSN signals is shown in Figure \ref{fig:ch5_min_SNR}. We select 5 representative waveforms from each explosion mechanism. For the magnetorotational explosion mechanism, we select 5 waveforms from the Scheidegger \emph{et al.} waveform catalogue. For the neutrino explosion mechanism, we select the 3 waveforms from M\"uller \emph{et al.}, and the s20 and s20s progenitor waveforms from Andresen \emph{et al.}. All of the waveforms are injected into aLIGO and AdVirgo noise recoloured to design sensitivity at the same GPS time and sky position. The distance of each signal is then changed to increase the network SNR values. 

To see the improvement in minimum SNR gained by using signal models created from three-dimensional waveforms instead of two-dimensional waveforms, we calculate the minimum SNR needed using the old signal models used in \cite{powell:16b}. The results are shown in the top two panels in Figure \ref{fig:ch5_min_SNR}. A signal is considered as being detected by SMEE if the signal verses noise log Bayes factor is larger than 10. We find with the old signal models that the average SNR needed to detected the magnetorotational waveforms is SNR $\sim12$, and the average SNR needed to detect the neutrino mechanism waveforms is SNR $\sim35$.  

To ensure the robustness of the result, three-dimensional waveforms that were not included in the PCs are used. As there are currently no more available three-dimensional waveforms that can be used for this test, the PCs are made again leaving out the signal that we want to measure the minimum SNR for. 7\,PCs are then used for each different signal model. The left out signals are injected in recoloured noise, and the distance of each signal is then changed to increase the network SNR values. 

The results are shown in the bottom two panels of Figure \ref{fig:ch5_min_SNR}. Updating SMEE to three-dimensional signal models has increased the sensitivity of SMEE for all types of CCSN waveforms used in this study. The average SNR needed for detection has now decreased to SNR $\sim9$ for the magnetorotational model, and the average SNR needed to detect neutrino mechanism waveforms has decreased to SNR $\sim27$. We expect the minimum SNR value to decrease further as more neutrino mechanism waveforms are simulated and more robust common features are identified.

\section{Analysis}
\label{sec:analysis}

As in previous sections, we use 24 hours of the detector noise from the Initial LIGO S5 science run \cite{2009RPPh...72g6901A} and the Virgo VSR2 run, which has been recoloured to match the expected design sensitivity of the detectors \cite{vallisneri:15, gossan:16}. We add simulated signals to the recoloured S5 data. The first is 928 injections of the magnetorotational Scheidegger \emph{et al.} waveform models Sch 5 and Sch 13. The second is 928 injections of the neutrino mechanism M\"uller \emph{et al.} waveform models L15 and N20. To demonstrate how SMEE can reject glitches we use the 1000 loudest background events found by cWB in the 24 hours of data used in this study.  

\begin{figure*}[!t]
\centering
\includegraphics[width=0.32\textwidth]{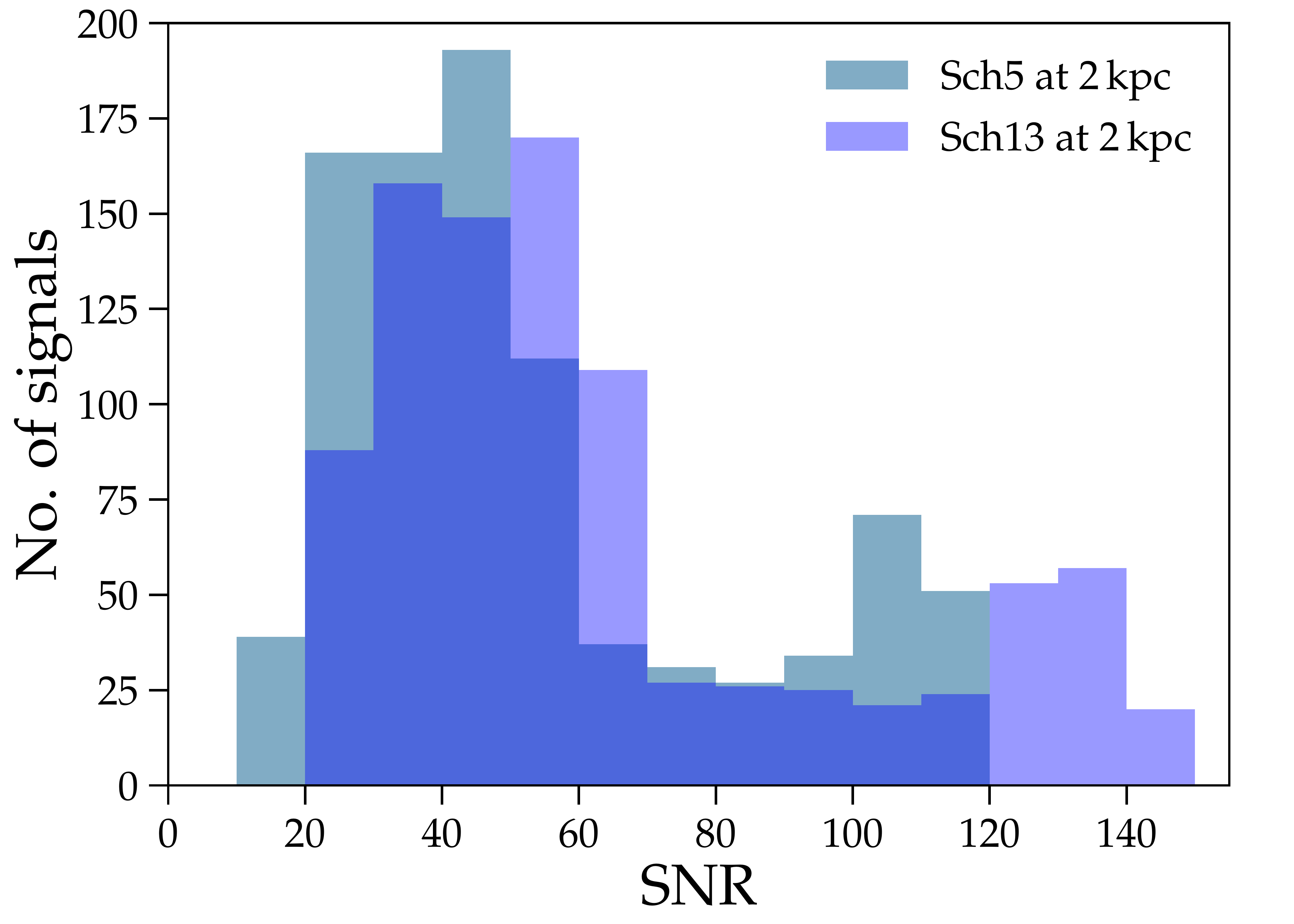}
\includegraphics[width=0.32\textwidth]{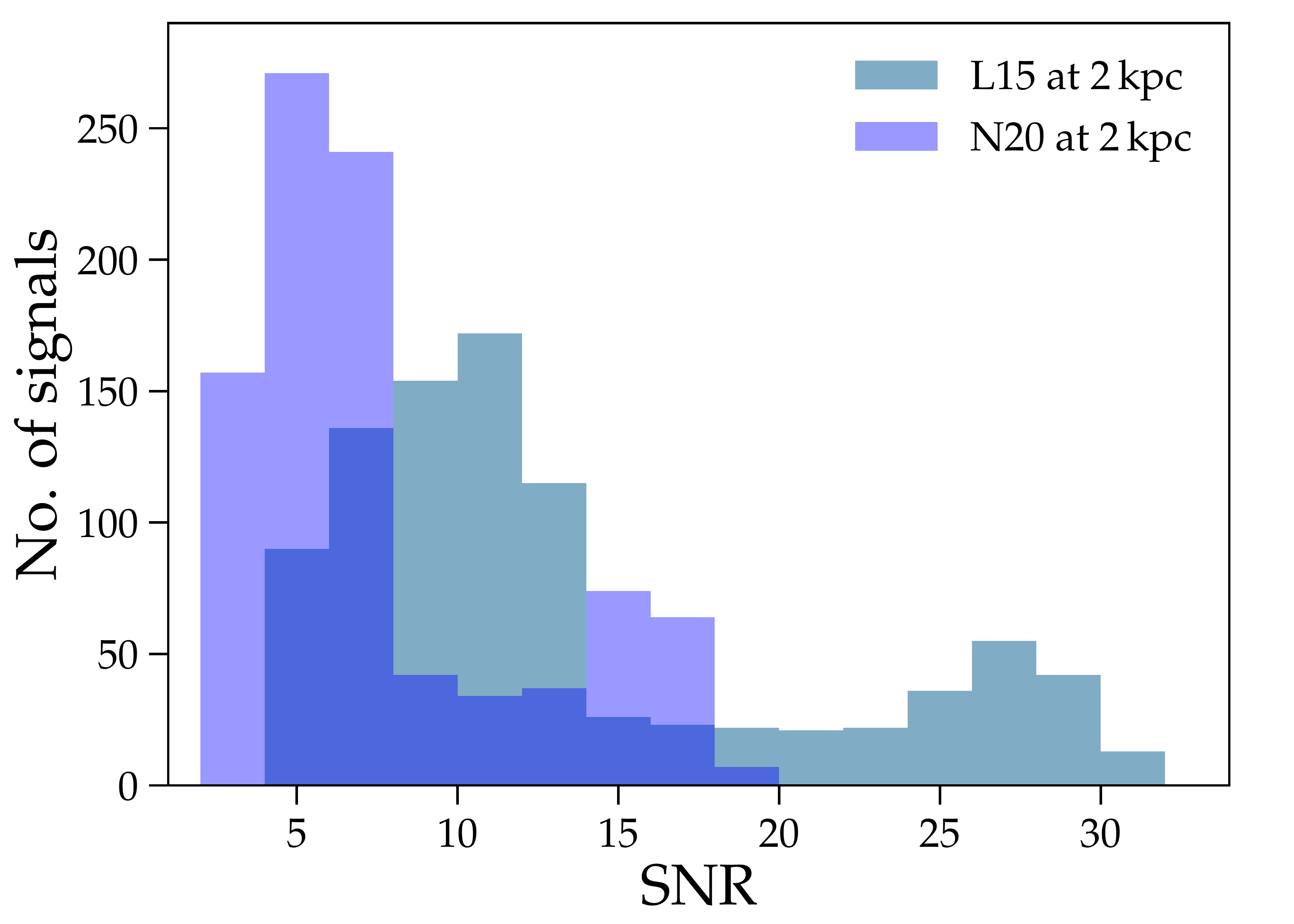}
\includegraphics[width=0.32\textwidth]{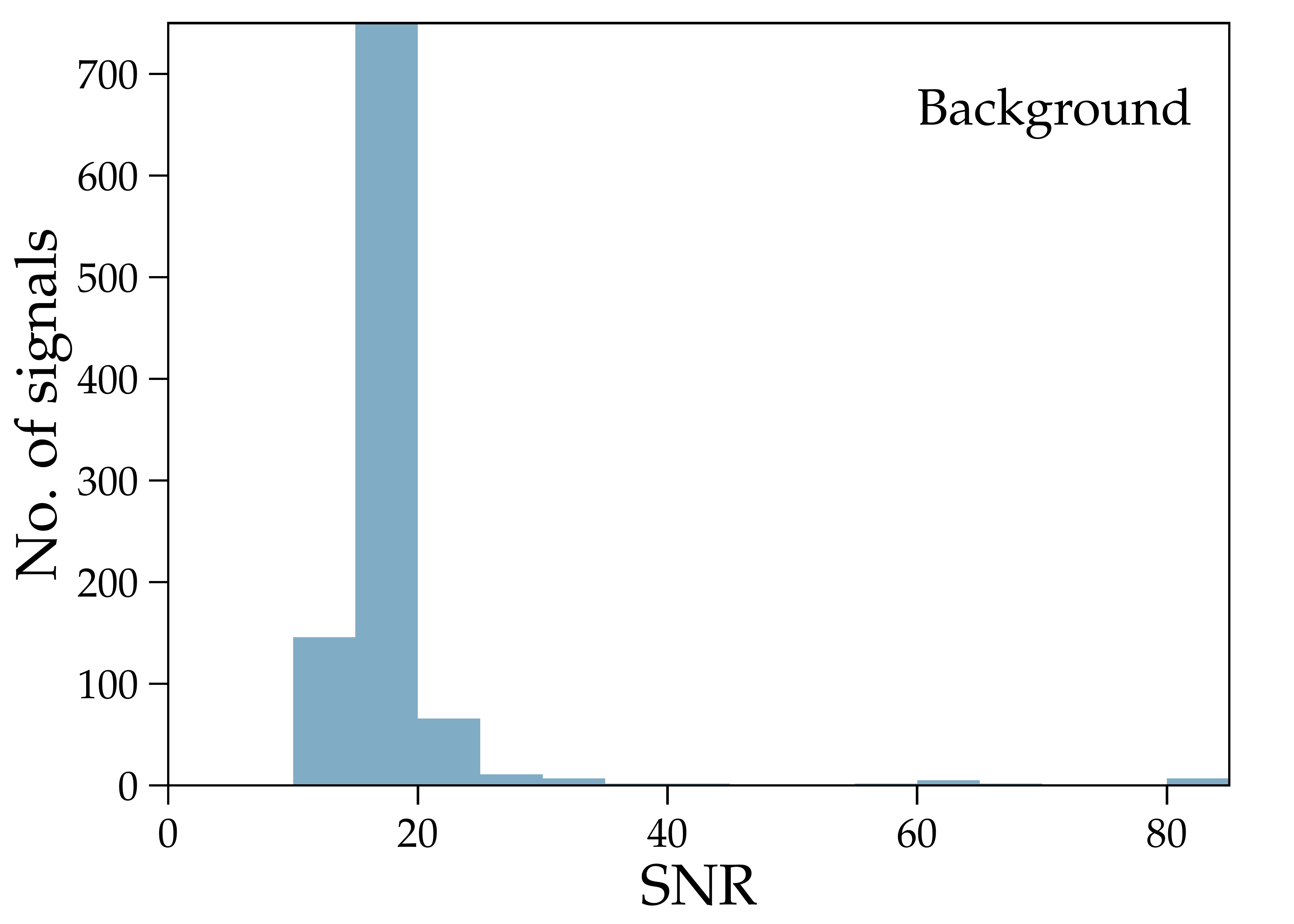}
\caption{(Left) The network SNR of the magnetorotational waveforms injected at $10\,$kpc. (Middle) The SNR of the neutrino mechanism waveforms injected at $2\,$kpc. (Right) The SNR of the background cWB triggers used in this study. The SNR depends on the antenna pattern at the time of the injection and the quality of the data. }
\label{fig:SNR_hists}
\end{figure*}

The magnetorotational waveforms are injected at a distance of $10\,$kpc, and the neutrino waveforms are injected at $2\,$kpc, as their gravitational-wave emission has a smaller amplitude, and the SNR needs to be large enough for the injections to be detected by SMEE. The SNR values of the magnetorotational and neutrino mechanism injections and the background triggers are shown in Figure \ref{fig:SNR_hists}. The SNR varies due to the antenna pattern and the quality of the data. The neutrino mechanism waveforms and the background triggers are all mainly below SNR 30. There are a few background triggers with an SNR as high as 80. The magnetorotational signals reach SNR values up to $\sim140$. 

As the distances used for the astrophysical signals are small enough that we would expect neutrinos to be detected from the source, we assume that the arrival time and sky position of the source are known. We treat the background triggers as potential candidates provided by the all-sky burst searches. 

\section{Results for signals included in model}
\label{sec:results}

\subsection{Glitch rejection}

Using both the magnetorotational and neutrino mechanism signal models we calculate signal vs. noise and coherent vs. incoherent Bayes factors for all of the signals used in this study. The aim is to reject signal candidates that are created by glitches in the detector before any further analysis of the signal. We start the analysis using signals that are included in the model, as these are the models that would be used for a real detection candidate, and it will demonstrate the current best possible results that can be obtained with SMEE. Including the injected signal in the model will produce better results than we would expect from a real CCSN signal candidate, as it is unlikely a real signal would be an exact match for one of the current available simulations. 

\begin{table}[ht!]
\begin{center}
\begin{tabular}{|c|ccc|ccc|}
\hline
 Injected signal & \multicolumn{3}{c|}{Neutrino model} & \multicolumn{3}{c|}{Magnetorotational} \\
  & \multicolumn{1}{c}{Signal} & \multicolumn{1}{c}{Glitch} & \multicolumn{1}{c|}{Noise}& \multicolumn{1}{c}{Signal}& \multicolumn{1}{c}{Glitch} & \multicolumn{1}{c|}{Noise}\\  \hline 
 L15 (neu)       & 686 & 16 & 226 & 35  & 0  & 893 \\
 N20 (neu)       & 190 & 14 & 724 & 22  & 0  & 906 \\ 
 Sch 5 (mag)& 246 & 0  & 682 & 928 & 0  & 0   \\
 Sch 13 (mag)& 20  & 0  & 908 & 928 & 0  & 0   \\ 
 glitch     & 0  & 163 & 837 & 0  & 97 & 903 \\  \hline
\end{tabular}
\caption{All signal candidates are classified as being signals, glitches or noise. None of the glitches are classified as signals. A small number of M\"uller neutrino waveforms are mis-classified as glitches. All Scheidegger magnetorotational waveforms are correctly classified as signals as they have a larger SNR.  }
\label{tab:classification}
\end{center}
\end{table}

The results are shown in Table \ref{tab:classification}. There are 686 L15 neutrino mechanism waveforms and 190 N20 neutrino mechanism waveforms correctly identified as signals when using the correct neutrino mechanism signal model. There are 226 L15 and 724 N20 signals that could not be distinguished from noise at 2\,kpc. A further 16 L15 and 14 N20 signals were incorrectly classified as being glitches. This occurs when the SNR in one or more of the detectors is below SNR 8. For the magnetorotational signals Sch 5 and Sch 13, all of them were detected and none were mis-classified as glitches when using the correct signal model. The improvement for the magnetorotational signals is due to their higher SNR at the larger distance of 10\,kpc. 

Using both the magnetorotational and neutrino mechanism signal models, all 1000 coincident glitches were rejected as potential signal candidates. The majority of glitches were classified as background noise. There are 163 and 97 glitches with a positive signal vs. noise Bayes factor, for the neutrino and magnetorotational models respectively, which failed the coherence test and are therefore classified as glitches.  

\subsection{Single detector glitch rejection}

The method implemented in the previous section for the rejection of glitches relies on multiple detectors being operational at the time that the signal candidate occurs. As the detectors do not have 100\% duty cycles it is possible that only one detector may be operational when a signal occurs. 

Here we outline how SMEE can be used to reject glitches in a single detector. We use the same signals and background triggers included in the previous section in Livingston detector data only. To determine if the signal candidate is a signal or glitch we create a signal model for glitches by applying PCA to the time series waveforms of the first 10 glitches that occur in the detector data. This method has been previously applied in detector characterization studies \cite{powell:15,powell:16}, but has never been applied as a method of rejecting single detector detection candidates. When a signal model for glitches has been created SMEE can calculate Bayes factors to determine if the signal candidates are glitches or real signals. 

\begin{table}[ht!]
\begin{center}
\begin{tabular}{|c|ccc|ccc|}
\hline
 Injected signal & \multicolumn{3}{c|}{Neutrino model} & \multicolumn{3}{c|}{Magnetorotational} \\
  & \multicolumn{1}{c}{Signal} & \multicolumn{1}{c}{Glitch} & \multicolumn{1}{c|}{Noise}& \multicolumn{1}{c}{Signal}& \multicolumn{1}{c}{Glitch} & \multicolumn{1}{c|}{Noise}\\  \hline 
 L15 (neu)       & 320 & 0   & 608 & 10  & 80  & 838 \\
 N20 (neu)       & 173 & 0   & 755 & 16  & 11  & 901 \\ 
 Sch 5 (mag)& 101 & 99  & 728 & 823 & 0   & 105 \\
 Sch 13 (mag)& 4   & 299 & 625 & 861 & 0   & 67  \\ 
 glitch     & 22  & 438 & 540 & 19  & 500 & 481 \\  \hline
\end{tabular}
\caption{The classification of signal candidates in Livingston detector data only. No astrophysical signals are mis-classified as glitches. There are $\sim 2\%$ of glitches mis-classified as signals.  }
\label{tab:singledet}
\end{center}
\end{table}

The results are shown in Table \ref{tab:singledet}. For the neutrino mechanism L15 signals, when using the correct neutrino mechanism signal model 320 of the signals are detected and none are mis-classified as glitches. For the neutrino mechanism N20 signals, when using the correct neutrino mechanism signal model 173 of the signals are detected and none are mis-classified as glitches. For the Sch 5 and Sch 13 magnetorotational signals there are 823 and 861 signals, respectively, that are detected and none are mis-classified as glitches. Approximately half of the glitches considered were detected by SMEE. Only $\sim 2\%$ of the glitches are mis-classified as being signals. 

\subsection{Determining the explosion mechanism}

Once it has been determined that the signal candidates were not glitches or background noise we determine the explosion mechanism of all of the candidates classified as signals. This is achieved by comparing the Bayes factor values for the two models. A signal is considered as being magnetorotational if $\mathrm{logB}_{MagRot-Neu} > 10$ and is considered to be neutrino mechanism if $\mathrm{logB}_{MagRot-Neu} < -10$. If the signal is between -10 and 10 then it is not possible to confidently determine what the explosion mechanism of the signal is. 

\begin{figure*}[!t]
\centering
\includegraphics[width=0.47\textwidth]{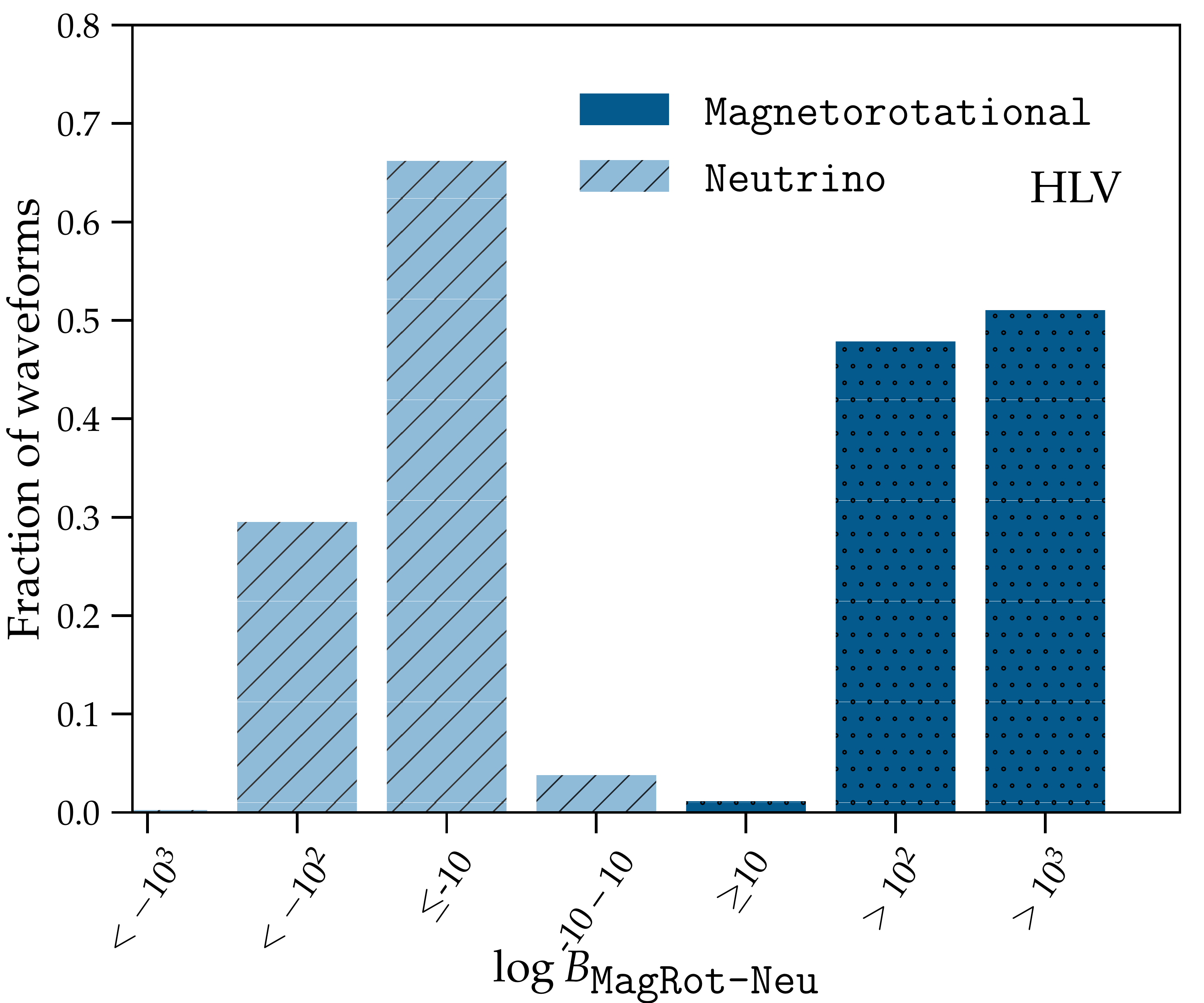}
\includegraphics[width=0.47\textwidth]{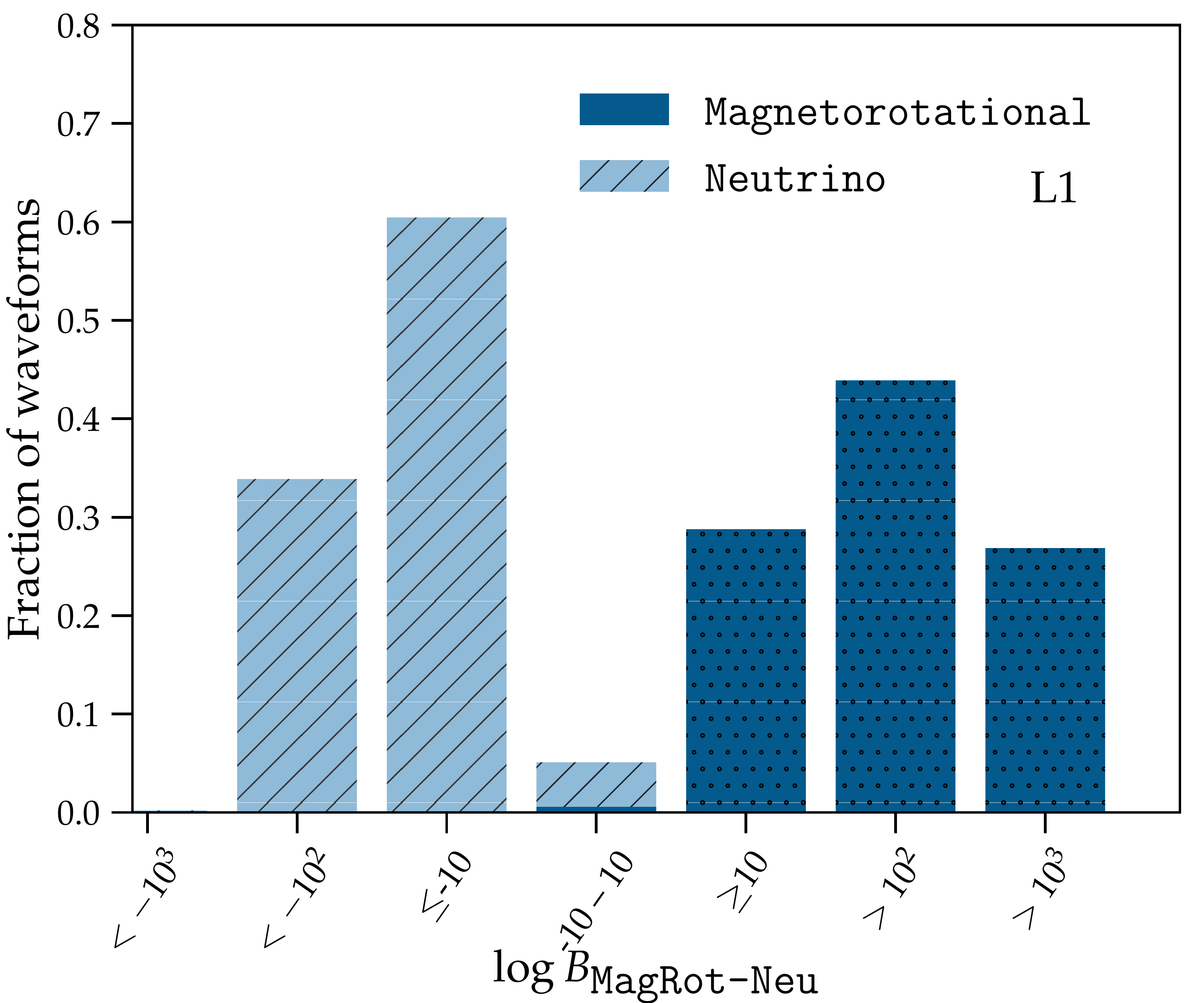}
\caption{The Bayes factors used to determine the explosion mechanism of the signal candidates. (Left) The results for a three detector network. (Right) The results when only the Livingston detector is considered. A signal is considered as being magnetorotational if $\mathrm{logB}_{MagRot-Neu} > 10$ and is considered to be neutrino mechanism if $\mathrm{logB}_{MagRot-Neu} < -10$.}
\label{fig:explosion_mech}
\end{figure*}

A histogram of the Bayes factors is shown in Figure \ref{fig:explosion_mech}. For the three detector network, all of the magnetorotational signals were correctly identified as being magnetorotational. SMEE could not determine the explosion mechanism for 39 of the neutrino mechanism signals. All the others were correctly identified as neutrino mechanism signals. 

For the single detector case, it was not possible to determine the explosion mechanism of 9 of the magnetorotational signals. All of the other injections were correctly identified as being magnetorotational mechanism signals. It was not possible to determine the explosion mechanism for 28 of the single detector neutrino mechanism signals. They are the signals with the lowest SNR values. All the others were correctly identified as being neutrino explosion mechanism signals

\section{Results for signals not included in model}
\label{sec:robust}
To test the robustness of the result, three-dimensional waveforms that were not included in the PCs are used. As there are currently no more available three-dimensional waveforms that can be used for this test, the PCs are made again leaving out signals from each of the explosion mechanisms. The waveforms left out of the signal models are the magnetorotational waveforms Sch 5, and the neutrino mechanism waveform L15. We use 7 PCs to represent each explosion mechanism. 
  
\subsection{Glitch rejection}
As in the previous sections, we inject the signals left out of the models, L15 and Sch 5, at 928 GPS times spread over the 24 hours of aLIGO and AdV recoloured detector noise. The results are shown in Table \ref{tab:robust_test}. A smaller number of signals are detected as more SNR is needed to detect signals that are not included in the signal model. The majority of detected injections are correctly classified as signals. When using the neutrino mechanism signal model no neutrino mechanism signals are mis-classified as glitches. When using the magnetorotational signal model there are 2 magnetorotational signals that are mis-classified as glitches. Over $\sim 80\%$ of the glitches are not detected by SMEE using either of the signal models.

\begin{table}[ht!]
\begin{center}
\begin{tabular}{|c|ccc|ccc|}
\hline
 Injected signal & \multicolumn{3}{c|}{Neutrino model} & \multicolumn{3}{c|}{Magnetorotational} \\
  & \multicolumn{1}{c}{Signal} & \multicolumn{1}{c}{Glitch} & \multicolumn{1}{c|}{Noise}& \multicolumn{1}{c}{Signal}& \multicolumn{1}{c}{Glitch} & \multicolumn{1}{c|}{Noise}\\  \hline 
 L15  (neu)      & 155 & 0   & 773 & 30  & 2   & 896 \\
 Sch 5 (mag) & 332 & 1   & 595 & 889 & 2   & 37  \\
 glitch     & 3   & 180 & 817 & 2   & 104 & 894 \\  \hline
\end{tabular}
\caption{The classification of two different signal types that were left out of the signal models to test robustness. A smaller number of signals are detected when the signal is not included in the model. }
\label{tab:robust_test}
\end{center}
\end{table}

\subsection{Single detector glitch rejection}

In this sub-section we determine how well SMEE can reject single detector glitches when the signal is not included in the model. As in the previous sub-section, the signals left out of the models are L15 and Sch 5. We then calculate Bayes factors to determine if the signal candidates are glitches or real signals. 

\begin{table}[ht!] 
\begin{center}
\begin{tabular}{|c|ccc|ccc|}
\hline
 Injected signal & \multicolumn{3}{c|}{Neutrino model} & \multicolumn{3}{c|}{Magnetorotational} \\
  & \multicolumn{1}{c}{Signal} & \multicolumn{1}{c}{Glitch} & \multicolumn{1}{c|}{Noise}& \multicolumn{1}{c}{Signal}& \multicolumn{1}{c}{Glitch} & \multicolumn{1}{c|}{Noise}\\  \hline 
 L15  (neu) & 153 & 34 & 741 & 51 & 45 & 832 \\
 Sch 5 (mag) & 188 & 72 & 668 & 629 & 1 & 298  \\
 glitch     & 1 & 384 & 615 & 1 & 171 & 828 \\  \hline
\end{tabular}
\caption{ The classification of different signals and glitches in a single detector when the injected signals are not included in the signal models. }
\label{tab:robust_test_liv}
\end{center}
\end{table}

The results are shown in Table \ref{tab:robust_test_liv}. For the neutrino mechanism L15 signals, when using the correct neutrino mechanism signal model 153 of the signals are detected and 34 are mis-classified as glitches. The number of neutrino mechanism waveforms mis-classified is larger when the signal is not included in the model. For the Sch 5 magnetorotational signals there are 741 signals that are detected and 1 is mis-classified as a glitch. Only 1 of the glitches are mis-classified as being a signal when using the neutrino mechanism model or the magnetorotational mechanism model.

\subsection{Determining the explosion mechanism}

\begin{figure*}[!t]
\centering
\includegraphics[width=\columnwidth]{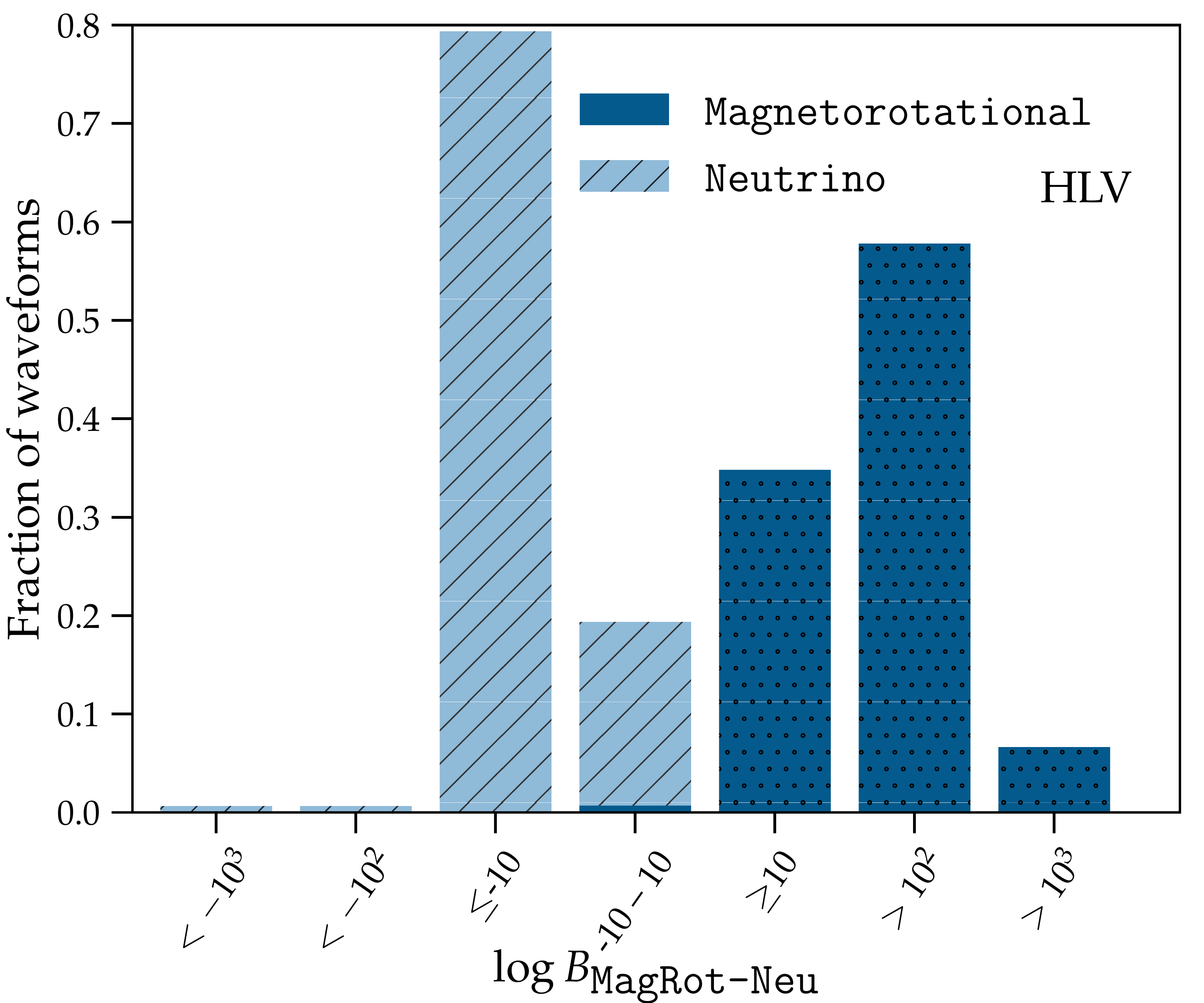}
\includegraphics[width=\columnwidth]{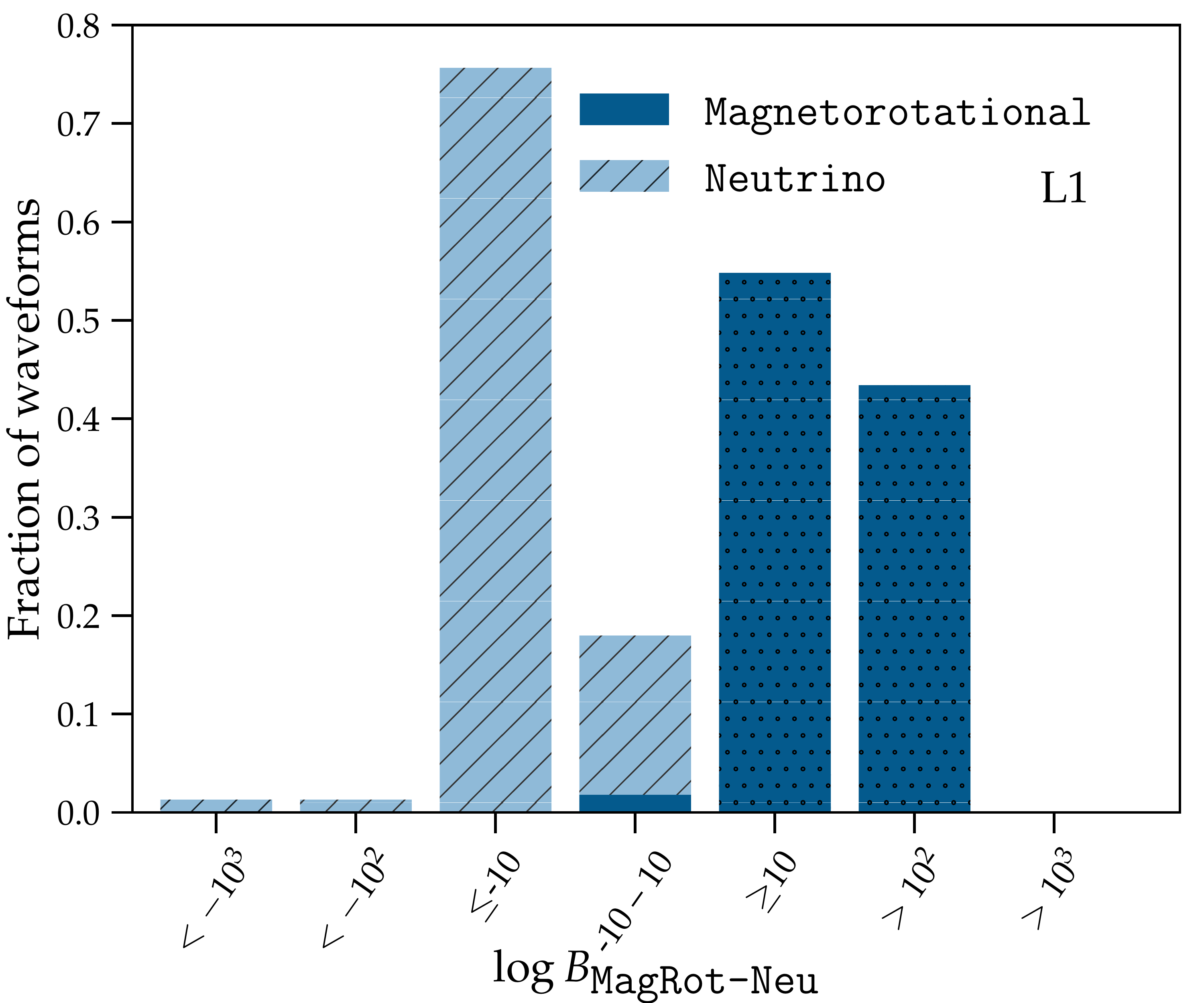}
\caption{The Bayes factors used to determine the explosion mechanism when two signals are left out of the signal models to test robustness. Left is the results for a three detector network. Right is the results for one LIGO detector only. A signal is considered as being magnetorotational if $\mathrm{logB}_{MagRot-Neu} > 10$ and is considered to be neutrino mechanism if $\mathrm{logB}_{MagRot-Neu} < -10$.}
\label{fig:explosion_mech_robust}
\end{figure*}

After the glitch rejection we determine the explosion mechanism for the remaining signal candidates in a single detector and a three detector network. As in the previous sections, a signal is considered as being magnetorotational if $\mathrm{logB}_{MagRot-Neu} > 10$ and is considered to be a neutrino mechanism signal if $\mathrm{logB}_{MagRot-Neu} < -10$. 

The results are shown in Figure \ref{fig:explosion_mech_robust}. For the three detector network, it was not possible to determine the explosion mechanism of 7 of the magnetorotational signals. All the other magnetorotational signals were correctly identified as being magnetorotational mechanism signals. It was not possible to determine the explosion mechanism of 30 of the neutrino mechanism signals. All other neutrino mechanism signals were correctly identified as being neutrino explosion mechanism signals. The number of waveforms for which the explosion mechanism could not be determined is similar to the previous results when the signal is included in the model. 

In the one detector network, it was not possible to distinguish the explosion mechanism for 10 magnetorotational waveforms and 75 neutrino waveforms. This is a larger number of neutrino mechanism signals than the previous results where the signal is included in the model. The explosion mechanism was correctly determined for all other waveforms detected in a single detector.  

\section{Discussion}
\label{sec:discussion}

During the advanced gravitational-wave detector observing runs, SMEE is used as a parameter estimation follow-up tool for potential CCSN signal candidates that are identified by the gravitational-wave searches, or from alerts sent by electromagnetic and neutrino detectors. In this paper, we update the signal models in SMEE to use three-dimensional neutrino mechanism and magnetorotational mechanism waveforms. Improvements to CCSN simulations have advanced rapidly in recent years, and using the latest available waveforms will maximise the potential for a gravitational-wave CCSN detection and measurement of the signals astrophysical parameters. A more robust result will be possible in the future as more three-dimensional CCSN simulations become available. The minimum SNR needed to detect the three-dimensional waveforms was improved by using signal models created from the three-dimensional waveforms.    

A real CCSN gravitational-wave signal candidate could potentially be a glitch in the detector noise. In this paper, we test the ability of SMEE to reject a gravitational-wave signal candidate if it is created by a glitch. It is shown that glitches can be eliminated as CCSN candidates by using Bayes factors to determine if the signal is coherent between all of the detectors being considered. Furthermore, we have outlined a new procedure for rejecting glitches when only one detector is operational. This could be extended further as a method for the detection of other types of gravitational-wave signals. 

Current searches for gravitational waves from CCSNe use a signal model that is generic and assumes no previous knowledge of what features should be expected in the signal \cite{gossan:16, 2016arXiv160501785A}. SMEE is a partially modelled method that assumes knowledge of the signal features. If the model is correct then we expect it to be more sensitive than the unmodelled searches. If the knowledge we are assuming about the signal is wrong then the unmodelled searches will have the better sensitivity. 

Currently it would not be possible for SMEE to replace the unmodelled supernova searches as an estimation of the significance of our detection would be too computationally expensive. This may be possible in the future with improvements to the speed of the algorithm. It is still important for us to show what parameter estimation methods can detect independently of the unmodelled searches, because the unmodelled searches only estimate the frequency, duration and amplitude of the signal. Methods designed to measure other astrophysical parameters need to be able to detect the signal to add further astrophysical information.   

In the future, we plan to apply PCA to spectrograms of the waveforms so that unreliable phase information will not be included in the signal models. As the CCSN rates are low for ground based advanced gravitational-wave detectors, it will be important to test the capability of SMEE for the next generation of gravitational-wave detectors.  

\begin{acknowledgments}
We thank John Vietch and Jonathan Gair for helpful comments on our work. We thank Sarah 
Gossan for discussions, extensive figure presentation help and for the creation of the recoloured 
noise used in this study. We thank the CCSN simulation community for making
their gravitational waveform predictions available for this
study. ISH and JP are supported by UK 
Science and Technology Facilities Council grants ST/L000946/1 and ST/L000946/1. JP is supported 
by the Australian Research Council Centre of Excellence for Gravitational Wave Discovery 
(OzGrav), through project number CE170100004.
The authors also gratefully acknowledge the support of the Scottish 
Universities Physics Alliance (SUPA). LIGO was constructed by the California Institute of Technology
and Massachusetts Institute of Technology with funding from the
National Science Foundation and operates under cooperative agreements
PHY-0107417 and PHY-0757058. 
\end{acknowledgments}


\bibliographystyle{apsrev}

\bibliography{bibfile}

\begin{thebibliography}{56}
\expandafter\ifx\csname natexlab\endcsname\relax\def\natexlab#1{#1}\fi
\expandafter\ifx\csname bibnamefont\endcsname\relax
  \def\bibnamefont#1{#1}\fi
\expandafter\ifx\csname bibfnamefont\endcsname\relax
  \def\bibfnamefont#1{#1}\fi
\expandafter\ifx\csname citenamefont\endcsname\relax
  \def\citenamefont#1{#1}\fi
\expandafter\ifx\csname url\endcsname\relax
  \def\url#1{\texttt{#1}}\fi
\expandafter\ifx\csname urlprefix\endcsname\relax\def\urlprefix{URL }\fi
\providecommand{\bibinfo}[2]{#2}
\providecommand{\eprint}[2][]{\url{#2}}

\bibitem[{\citenamefont{{The LIGO Scientific Collaboration}
  et~al.}(2015)\citenamefont{{The LIGO Scientific Collaboration}, {Aasi},
  {Abbott}, {Abbott}, and et~al.}}]{aLIGO}
\bibinfo{author}{\bibnamefont{{The LIGO Scientific Collaboration}}},
  \bibinfo{author}{\bibfnamefont{J.}~\bibnamefont{{Aasi}}},
  \bibinfo{author}{\bibfnamefont{B.~P.} \bibnamefont{{Abbott}}},
  \bibinfo{author}{\bibfnamefont{R.}~\bibnamefont{{Abbott}}}, \bibnamefont{and}
  \bibinfo{author}{\bibnamefont{et~al.}}, \bibinfo{journal}{\cqg}
  \textbf{\bibinfo{volume}{32}}, \bibinfo{eid}{074001} (\bibinfo{year}{2015}),
  \eprint{1411.4547}.

\bibitem[{\citenamefont{Abbott et~al.}(2016{\natexlab{a}})}]{GW150914}
\bibinfo{author}{\bibfnamefont{B.~P.} \bibnamefont{Abbott}}
  \bibnamefont{et~al.} (\bibinfo{collaboration}{LIGO Scientific Collaboration
  and Virgo Collaboration}), \bibinfo{journal}{Phys. Rev. Lett.}
  \textbf{\bibinfo{volume}{116}}, \bibinfo{pages}{061102}
  (\bibinfo{year}{2016}{\natexlab{a}}),
  \urlprefix\url{http://link.aps.org/doi/10.1103/PhysRevLett.116.061102}.

\bibitem[{\citenamefont{Abbott et~al.}(2016{\natexlab{b}})}]{GW151226}
\bibinfo{author}{\bibfnamefont{B.~P.} \bibnamefont{Abbott}}
  \bibnamefont{et~al.} (\bibinfo{collaboration}{LIGO Scientific Collaboration
  and Virgo Collaboration}), \bibinfo{journal}{Phys. Rev. Lett.}
  \textbf{\bibinfo{volume}{116}}, \bibinfo{pages}{241103}
  (\bibinfo{year}{2016}{\natexlab{b}}),
  \urlprefix\url{http://link.aps.org/doi/10.1103/PhysRevLett.116.241103}.

\bibitem[{\citenamefont{{Abbott}
  et~al.}(2016{\natexlab{a}})\citenamefont{{Abbott}, {Abbott}, {Abbott},
  {Abernathy}, {Acernese}, {Ackley}, {Adams}, {Adams}, {Addesso}, {Adhikari}
  et~al.}}]{o1cbcsearch}
\bibinfo{author}{\bibfnamefont{B.~P.} \bibnamefont{{Abbott}}},
  \bibinfo{author}{\bibfnamefont{R.}~\bibnamefont{{Abbott}}},
  \bibinfo{author}{\bibfnamefont{T.~D.} \bibnamefont{{Abbott}}},
  \bibinfo{author}{\bibfnamefont{M.~R.} \bibnamefont{{Abernathy}}},
  \bibinfo{author}{\bibfnamefont{F.}~\bibnamefont{{Acernese}}},
  \bibinfo{author}{\bibfnamefont{K.}~\bibnamefont{{Ackley}}},
  \bibinfo{author}{\bibfnamefont{C.}~\bibnamefont{{Adams}}},
  \bibinfo{author}{\bibfnamefont{T.}~\bibnamefont{{Adams}}},
  \bibinfo{author}{\bibfnamefont{P.}~\bibnamefont{{Addesso}}},
  \bibinfo{author}{\bibfnamefont{R.~X.} \bibnamefont{{Adhikari}}},
  \bibnamefont{et~al.}, \bibinfo{journal}{Physical Review X}
  \textbf{\bibinfo{volume}{6}}, \bibinfo{eid}{041015}
  (\bibinfo{year}{2016}{\natexlab{a}}), \eprint{1606.04856}.

\bibitem[{\citenamefont{{Acernese} and et~al.}(2015)}]{AdVirgo}
\bibinfo{author}{\bibfnamefont{F.}~\bibnamefont{{Acernese}}} \bibnamefont{and}
  \bibinfo{author}{\bibnamefont{et~al.}}, \bibinfo{journal}{\cqg}
  \textbf{\bibinfo{volume}{32}}, \bibinfo{eid}{024001} (\bibinfo{year}{2015}),
  \eprint{1408.3978}.

\bibitem[{\citenamefont{{Ott} et~al.}(2004)\citenamefont{{Ott}, {Burrows},
  {Livne}, and {Walder}}}]{ott:04}
\bibinfo{author}{\bibfnamefont{C.~D.} \bibnamefont{{Ott}}},
  \bibinfo{author}{\bibfnamefont{A.}~\bibnamefont{{Burrows}}},
  \bibinfo{author}{\bibfnamefont{E.}~\bibnamefont{{Livne}}}, \bibnamefont{and}
  \bibinfo{author}{\bibfnamefont{R.}~\bibnamefont{{Walder}}},
  \bibinfo{journal}{\apj} \textbf{\bibinfo{volume}{600}}, \bibinfo{pages}{834}
  (\bibinfo{year}{2004}), \eprint{astro-ph/0307472}.

\bibitem[{\citenamefont{{Abbott}
  et~al.}(2016{\natexlab{b}})}]{2016arXiv160501785A}
\bibinfo{author}{\bibfnamefont{B.~P.} \bibnamefont{{Abbott}}}
  \bibnamefont{et~al.} (\bibinfo{collaboration}{LIGO Scientific Collaboration
  and Virgo Collaboration}), \bibinfo{journal}{Phys. Rev. D}
  \textbf{\bibinfo{volume}{94}}, \bibinfo{pages}{102001}
  (\bibinfo{year}{2016}{\natexlab{b}}),
  \urlprefix\url{http://link.aps.org/doi/10.1103/PhysRevD.94.102001}.

\bibitem[{\citenamefont{{Gossan} et~al.}(2016)\citenamefont{{Gossan}, {Sutton},
  {Stuver}, {Zanolin}, {Gill}, and {Ott}}}]{gossan:16}
\bibinfo{author}{\bibfnamefont{S.~E.} \bibnamefont{{Gossan}}},
  \bibinfo{author}{\bibfnamefont{P.}~\bibnamefont{{Sutton}}},
  \bibinfo{author}{\bibfnamefont{A.}~\bibnamefont{{Stuver}}},
  \bibinfo{author}{\bibfnamefont{M.}~\bibnamefont{{Zanolin}}},
  \bibinfo{author}{\bibfnamefont{K.}~\bibnamefont{{Gill}}}, \bibnamefont{and}
  \bibinfo{author}{\bibfnamefont{C.~D.} \bibnamefont{{Ott}}},
  \bibinfo{journal}{\prd} \textbf{\bibinfo{volume}{93}}, \bibinfo{eid}{042002}
  (\bibinfo{year}{2016}).

\bibitem[{\citenamefont{{Powell} et~al.}(2016)\citenamefont{{Powell}, {Gossan},
  {Logue}, and {Heng}}}]{powell:16b}
\bibinfo{author}{\bibfnamefont{J.}~\bibnamefont{{Powell}}},
  \bibinfo{author}{\bibfnamefont{S.~E.} \bibnamefont{{Gossan}}},
  \bibinfo{author}{\bibfnamefont{J.}~\bibnamefont{{Logue}}}, \bibnamefont{and}
  \bibinfo{author}{\bibfnamefont{I.~S.} \bibnamefont{{Heng}}},
  \bibinfo{journal}{\prd} \textbf{\bibinfo{volume}{94}}, \bibinfo{eid}{123012}
  (\bibinfo{year}{2016}), \eprint{1610.05573}.

\bibitem[{\citenamefont{{Logue} et~al.}(2012)\citenamefont{{Logue}, {Ott},
  {Heng}, {Kalmus}, and {Scargill}}}]{logue:12}
\bibinfo{author}{\bibfnamefont{J.}~\bibnamefont{{Logue}}},
  \bibinfo{author}{\bibfnamefont{C.~D.} \bibnamefont{{Ott}}},
  \bibinfo{author}{\bibfnamefont{I.~S.} \bibnamefont{{Heng}}},
  \bibinfo{author}{\bibfnamefont{P.}~\bibnamefont{{Kalmus}}}, \bibnamefont{and}
  \bibinfo{author}{\bibfnamefont{J.}~\bibnamefont{{Scargill}}},
  \bibinfo{journal}{\prd} \textbf{\bibinfo{volume}{86}}, \bibinfo{eid}{044023}
  (\bibinfo{year}{2012}).

\bibitem[{\citenamefont{{Baron} and {Cooperstein}}(1990)}]{baron:90}
\bibinfo{author}{\bibfnamefont{E.}~\bibnamefont{{Baron}}} \bibnamefont{and}
  \bibinfo{author}{\bibfnamefont{J.}~\bibnamefont{{Cooperstein}}},
  \bibinfo{journal}{\apj} \textbf{\bibinfo{volume}{353}}, \bibinfo{pages}{597}
  (\bibinfo{year}{1990}).

\bibitem[{\citenamefont{{Bethe}}(1990)}]{bethe:90}
\bibinfo{author}{\bibfnamefont{H.~A.} \bibnamefont{{Bethe}}},
  \bibinfo{journal}{Reviews of Modern Physics} \textbf{\bibinfo{volume}{62}},
  \bibinfo{pages}{801} (\bibinfo{year}{1990}).

\bibitem[{\citenamefont{{O'Connor} and {Ott}}(2011)}]{oconnor:11}
\bibinfo{author}{\bibfnamefont{E.}~\bibnamefont{{O'Connor}}} \bibnamefont{and}
  \bibinfo{author}{\bibfnamefont{C.~D.} \bibnamefont{{Ott}}},
  \bibinfo{journal}{\apj} \textbf{\bibinfo{volume}{730}}, \bibinfo{eid}{70}
  (\bibinfo{year}{2011}), \eprint{1010.5550}.

\bibitem[{\citenamefont{{Skilling}}(2004)}]{skilling:04}
\bibinfo{author}{\bibfnamefont{J.}~\bibnamefont{{Skilling}}}, in
  \emph{\bibinfo{booktitle}{American Institute of Physics Conference Series}},
  edited by \bibinfo{editor}{\bibfnamefont{R.}~\bibnamefont{{Fischer}}},
  \bibinfo{editor}{\bibfnamefont{R.}~\bibnamefont{{Preuss}}}, \bibnamefont{and}
  \bibinfo{editor}{\bibfnamefont{U.~V.} \bibnamefont{{Toussaint}}}
  (\bibinfo{year}{2004}), vol. \bibinfo{volume}{735} of
  \emph{\bibinfo{series}{American Institute of Physics Conference Series}}, pp.
  \bibinfo{pages}{395--405}.

\bibitem[{\citenamefont{Sivia}(1996)}]{sivia:96}
\bibinfo{author}{\bibfnamefont{D.~S.} \bibnamefont{Sivia}},
  \emph{\bibinfo{title}{Data Analysis, A Bayesian Tutorial}}
  (\bibinfo{publisher}{Oxford}, \bibinfo{year}{1996}).

\bibitem[{\citenamefont{{Summerscales}
  et~al.}(2008)\citenamefont{{Summerscales}, {Burrows}, {Finn}, and
  {Ott}}}]{summerscales:08}
\bibinfo{author}{\bibfnamefont{T.~Z.} \bibnamefont{{Summerscales}}},
  \bibinfo{author}{\bibfnamefont{A.}~\bibnamefont{{Burrows}}},
  \bibinfo{author}{\bibfnamefont{L.~S.} \bibnamefont{{Finn}}},
  \bibnamefont{and} \bibinfo{author}{\bibfnamefont{C.~D.} \bibnamefont{{Ott}}},
  \bibinfo{journal}{\apj} \textbf{\bibinfo{volume}{678}},
  \bibinfo{eid}{1142-1157} (\bibinfo{year}{2008}), \eprint{0704.2157}.

\bibitem[{\citenamefont{Brady and Ray-Majumder}(2004)}]{brady:04}
\bibinfo{author}{\bibfnamefont{P.~R.} \bibnamefont{Brady}} \bibnamefont{and}
  \bibinfo{author}{\bibfnamefont{S.}~\bibnamefont{Ray-Majumder}},
  \bibinfo{journal}{Classical and Quantum Gravity}
  \textbf{\bibinfo{volume}{21}}, \bibinfo{pages}{S1839} (\bibinfo{year}{2004}),
  \urlprefix\url{http://stacks.iop.org/0264-9381/21/i=20/a=027}.

\bibitem[{\citenamefont{{Heng}}(2009)}]{heng:09}
\bibinfo{author}{\bibfnamefont{I.~S.} \bibnamefont{{Heng}}},
  \bibinfo{journal}{\cqg} \textbf{\bibinfo{volume}{26}}, \bibinfo{eid}{105005}
  (\bibinfo{year}{2009}), \eprint{0810.5707}.

\bibitem[{\citenamefont{{Dimmelmeier} et~al.}(2008)\citenamefont{{Dimmelmeier},
  {Ott}, {Marek}, and {Janka}}}]{dimmelmeier:08}
\bibinfo{author}{\bibfnamefont{H.}~\bibnamefont{{Dimmelmeier}}},
  \bibinfo{author}{\bibfnamefont{C.~D.} \bibnamefont{{Ott}}},
  \bibinfo{author}{\bibfnamefont{A.}~\bibnamefont{{Marek}}}, \bibnamefont{and}
  \bibinfo{author}{\bibfnamefont{H.-T.} \bibnamefont{{Janka}}},
  \bibinfo{journal}{\prd} \textbf{\bibinfo{volume}{78}},
  \bibinfo{pages}{064056} (\bibinfo{year}{2008}).

\bibitem[{\citenamefont{{R{\"o}ver} et~al.}(2009)\citenamefont{{R{\"o}ver},
  {Bizouard}, {Christensen}, {Dimmelmeier}, {Heng}, and {Meyer}}}]{roever:09}
\bibinfo{author}{\bibfnamefont{C.}~\bibnamefont{{R{\"o}ver}}},
  \bibinfo{author}{\bibfnamefont{M.-A.} \bibnamefont{{Bizouard}}},
  \bibinfo{author}{\bibfnamefont{N.}~\bibnamefont{{Christensen}}},
  \bibinfo{author}{\bibfnamefont{H.}~\bibnamefont{{Dimmelmeier}}},
  \bibinfo{author}{\bibfnamefont{I.~S.} \bibnamefont{{Heng}}},
  \bibnamefont{and} \bibinfo{author}{\bibfnamefont{R.}~\bibnamefont{{Meyer}}},
  \bibinfo{journal}{\prd} \textbf{\bibinfo{volume}{80}}, \bibinfo{eid}{102004}
  (\bibinfo{year}{2009}), \eprint{0909.1093}.

\bibitem[{\citenamefont{{Cannon} et~al.}(2010)\citenamefont{{Cannon},
  {Chapman}, {Hanna}, {Keppel}, {Searle}, and {Weinstein}}}]{cannon:10}
\bibinfo{author}{\bibfnamefont{K.}~\bibnamefont{{Cannon}}},
  \bibinfo{author}{\bibfnamefont{A.}~\bibnamefont{{Chapman}}},
  \bibinfo{author}{\bibfnamefont{C.}~\bibnamefont{{Hanna}}},
  \bibinfo{author}{\bibfnamefont{D.}~\bibnamefont{{Keppel}}},
  \bibinfo{author}{\bibfnamefont{A.~C.} \bibnamefont{{Searle}}},
  \bibnamefont{and} \bibinfo{author}{\bibfnamefont{A.~J.}
  \bibnamefont{{Weinstein}}}, \bibinfo{journal}{\prd}
  \textbf{\bibinfo{volume}{82}}, \bibinfo{eid}{044025} (\bibinfo{year}{2010}),
  \eprint{1005.0012}.

\bibitem[{\citenamefont{{Smith} et~al.}(2013)\citenamefont{{Smith}, {Cannon},
  {Hanna}, {Keppel}, and {Mandel}}}]{smith:13}
\bibinfo{author}{\bibfnamefont{R.~J.~E.} \bibnamefont{{Smith}}},
  \bibinfo{author}{\bibfnamefont{K.}~\bibnamefont{{Cannon}}},
  \bibinfo{author}{\bibfnamefont{C.}~\bibnamefont{{Hanna}}},
  \bibinfo{author}{\bibfnamefont{D.}~\bibnamefont{{Keppel}}}, \bibnamefont{and}
  \bibinfo{author}{\bibfnamefont{I.}~\bibnamefont{{Mandel}}},
  \bibinfo{journal}{\prd} \textbf{\bibinfo{volume}{87}}, \bibinfo{eid}{122002}
  (\bibinfo{year}{2013}), \eprint{1211.1254}.

\bibitem[{\citenamefont{Pürrer}(2014)}]{0264-9381-31-19-195010}
\bibinfo{author}{\bibfnamefont{M.}~\bibnamefont{Pürrer}},
  \bibinfo{journal}{Classical and Quantum Gravity}
  \textbf{\bibinfo{volume}{31}}, \bibinfo{pages}{195010}
  (\bibinfo{year}{2014}),
  \urlprefix\url{http://stacks.iop.org/0264-9381/31/i=19/a=195010}.

\bibitem[{\citenamefont{Blackman et~al.}(2015)\citenamefont{Blackman, Field,
  Galley, Szil\'agyi, Scheel, Tiglio, and Hemberger}}]{PhysRevLett.115.121102}
\bibinfo{author}{\bibfnamefont{J.}~\bibnamefont{Blackman}},
  \bibinfo{author}{\bibfnamefont{S.~E.} \bibnamefont{Field}},
  \bibinfo{author}{\bibfnamefont{C.~R.} \bibnamefont{Galley}},
  \bibinfo{author}{\bibfnamefont{B.}~\bibnamefont{Szil\'agyi}},
  \bibinfo{author}{\bibfnamefont{M.~A.} \bibnamefont{Scheel}},
  \bibinfo{author}{\bibfnamefont{M.}~\bibnamefont{Tiglio}}, \bibnamefont{and}
  \bibinfo{author}{\bibfnamefont{D.~A.} \bibnamefont{Hemberger}},
  \bibinfo{journal}{Phys. Rev. Lett.} \textbf{\bibinfo{volume}{115}},
  \bibinfo{pages}{121102} (\bibinfo{year}{2015}),
  \urlprefix\url{http://link.aps.org/doi/10.1103/PhysRevLett.115.121102}.

\bibitem[{\citenamefont{Clark et~al.}(2016)\citenamefont{Clark, Bauswein,
  Stergioulas, and Shoemaker}}]{0264-9381-33-8-085003}
\bibinfo{author}{\bibfnamefont{J.~A.} \bibnamefont{Clark}},
  \bibinfo{author}{\bibfnamefont{A.}~\bibnamefont{Bauswein}},
  \bibinfo{author}{\bibfnamefont{N.}~\bibnamefont{Stergioulas}},
  \bibnamefont{and}
  \bibinfo{author}{\bibfnamefont{D.}~\bibnamefont{Shoemaker}},
  \bibinfo{journal}{Classical and Quantum Gravity}
  \textbf{\bibinfo{volume}{33}}, \bibinfo{pages}{085003}
  (\bibinfo{year}{2016}),
  \urlprefix\url{http://stacks.iop.org/0264-9381/33/i=8/a=085003}.

\bibitem[{\citenamefont{{Powell} et~al.}(2015)\citenamefont{{Powell},
  {Trifir{\`o}}, {Cuoco}, {Heng}, and {Cavagli{\`a}}}}]{powell:15}
\bibinfo{author}{\bibfnamefont{J.}~\bibnamefont{{Powell}}},
  \bibinfo{author}{\bibfnamefont{D.}~\bibnamefont{{Trifir{\`o}}}},
  \bibinfo{author}{\bibfnamefont{E.}~\bibnamefont{{Cuoco}}},
  \bibinfo{author}{\bibfnamefont{I.~S.} \bibnamefont{{Heng}}},
  \bibnamefont{and}
  \bibinfo{author}{\bibfnamefont{M.}~\bibnamefont{{Cavagli{\`a}}}},
  \bibinfo{journal}{\cqg} \textbf{\bibinfo{volume}{32}}, \bibinfo{eid}{215012}
  (\bibinfo{year}{2015}), \eprint{1505.01299}.

\bibitem[{\citenamefont{{Powell} et~al.}(2017)\citenamefont{{Powell},
  {Torres-Forn{\'e}}, {Lynch}, {Trifir{\`o}}, {Cuoco}, {Cavagli{\`a}}, {Heng},
  and {Font}}}]{powell:16}
\bibinfo{author}{\bibfnamefont{J.}~\bibnamefont{{Powell}}},
  \bibinfo{author}{\bibfnamefont{A.}~\bibnamefont{{Torres-Forn{\'e}}}},
  \bibinfo{author}{\bibfnamefont{R.}~\bibnamefont{{Lynch}}},
  \bibinfo{author}{\bibfnamefont{D.}~\bibnamefont{{Trifir{\`o}}}},
  \bibinfo{author}{\bibfnamefont{E.}~\bibnamefont{{Cuoco}}},
  \bibinfo{author}{\bibfnamefont{M.}~\bibnamefont{{Cavagli{\`a}}}},
  \bibinfo{author}{\bibfnamefont{I.~S.} \bibnamefont{{Heng}}},
  \bibnamefont{and} \bibinfo{author}{\bibfnamefont{J.~A.}
  \bibnamefont{{Font}}}, \bibinfo{journal}{Classical and Quantum Gravity}
  \textbf{\bibinfo{volume}{34}}, \bibinfo{eid}{034002} (\bibinfo{year}{2017}),
  \eprint{1609.06262}.

\bibitem[{\citenamefont{{Murphy} et~al.}(2009)\citenamefont{{Murphy}, {Ott},
  and {Burrows}}}]{murphy:09}
\bibinfo{author}{\bibfnamefont{J.~W.} \bibnamefont{{Murphy}}},
  \bibinfo{author}{\bibfnamefont{C.~D.} \bibnamefont{{Ott}}}, \bibnamefont{and}
  \bibinfo{author}{\bibfnamefont{A.}~\bibnamefont{{Burrows}}},
  \bibinfo{journal}{\apj} \textbf{\bibinfo{volume}{707}}, \bibinfo{pages}{1173}
  (\bibinfo{year}{2009}), \eprint{0907.4762}.

\bibitem[{\citenamefont{{Burrows} et~al.}(2006)\citenamefont{{Burrows},
  {Livne}, {Dessart}, {Ott}, and {Murphy}}}]{burrows:06}
\bibinfo{author}{\bibfnamefont{A.}~\bibnamefont{{Burrows}}},
  \bibinfo{author}{\bibfnamefont{E.}~\bibnamefont{{Livne}}},
  \bibinfo{author}{\bibfnamefont{L.}~\bibnamefont{{Dessart}}},
  \bibinfo{author}{\bibfnamefont{C.~D.} \bibnamefont{{Ott}}}, \bibnamefont{and}
  \bibinfo{author}{\bibfnamefont{J.}~\bibnamefont{{Murphy}}},
  \bibinfo{journal}{\apj} \textbf{\bibinfo{volume}{640}}, \bibinfo{pages}{878}
  (\bibinfo{year}{2006}), \eprint{astro-ph/0510687}.

\bibitem[{\citenamefont{{Kuroda} et~al.}(2016)\citenamefont{{Kuroda}, {Kotake},
  and {Takiwaki}}}]{kuroda:16}
\bibinfo{author}{\bibfnamefont{T.}~\bibnamefont{{Kuroda}}},
  \bibinfo{author}{\bibfnamefont{K.}~\bibnamefont{{Kotake}}}, \bibnamefont{and}
  \bibinfo{author}{\bibfnamefont{T.}~\bibnamefont{{Takiwaki}}},
  \bibinfo{journal}{\apjl} \textbf{\bibinfo{volume}{829}}, \bibinfo{eid}{L14}
  (\bibinfo{year}{2016}), \eprint{1605.09215}.

\bibitem[{\citenamefont{{M{\"u}ller} et~al.}(2012)\citenamefont{{M{\"u}ller},
  {Janka}, and {Wongwathanarat}}}]{mueller:e12}
\bibinfo{author}{\bibfnamefont{E.}~\bibnamefont{{M{\"u}ller}}},
  \bibinfo{author}{\bibfnamefont{H.-T.} \bibnamefont{{Janka}}},
  \bibnamefont{and}
  \bibinfo{author}{\bibfnamefont{A.}~\bibnamefont{{Wongwathanarat}}},
  \bibinfo{journal}{\aap} \textbf{\bibinfo{volume}{537}}, \bibinfo{eid}{A63}
  (\bibinfo{year}{2012}), \eprint{1106.6301}.

\bibitem[{\citenamefont{{Scheidegger} et~al.}(2010)\citenamefont{{Scheidegger},
  {K{\"a}ppeli}, {Whitehouse}, {Fischer}, and
  {Liebend{\"o}rfer}}}]{scheidegger:10b}
\bibinfo{author}{\bibfnamefont{S.}~\bibnamefont{{Scheidegger}}},
  \bibinfo{author}{\bibfnamefont{R.}~\bibnamefont{{K{\"a}ppeli}}},
  \bibinfo{author}{\bibfnamefont{S.~C.} \bibnamefont{{Whitehouse}}},
  \bibinfo{author}{\bibfnamefont{T.}~\bibnamefont{{Fischer}}},
  \bibnamefont{and}
  \bibinfo{author}{\bibfnamefont{M.}~\bibnamefont{{Liebend{\"o}rfer}}},
  \bibinfo{journal}{\aap} \textbf{\bibinfo{volume}{514}}, \bibinfo{pages}{A51}
  (\bibinfo{year}{2010}).

\bibitem[{\citenamefont{{Andresen} et~al.}(2017)\citenamefont{{Andresen},
  {M{\"u}ller}, {M{\"u}ller}, and {Janka}}}]{andresen:16}
\bibinfo{author}{\bibfnamefont{H.}~\bibnamefont{{Andresen}}},
  \bibinfo{author}{\bibfnamefont{B.}~\bibnamefont{{M{\"u}ller}}},
  \bibinfo{author}{\bibfnamefont{E.}~\bibnamefont{{M{\"u}ller}}},
  \bibnamefont{and} \bibinfo{author}{\bibfnamefont{H.-T.}
  \bibnamefont{{Janka}}}, \bibinfo{journal}{\mnras}
  \textbf{\bibinfo{volume}{468}}, \bibinfo{pages}{2032} (\bibinfo{year}{2017}),
  \eprint{1607.05199}.

\bibitem[{\citenamefont{{Janka}}(2012)}]{janka:12a}
\bibinfo{author}{\bibfnamefont{H.-T.} \bibnamefont{{Janka}}},
  \bibinfo{journal}{Ann. Rev. Nuc. Par. Sci.} \textbf{\bibinfo{volume}{62}},
  \bibinfo{pages}{407} (\bibinfo{year}{2012}), \eprint{1206.2503}.

\bibitem[{\citenamefont{Burrows}(2013)}]{burrows:13}
\bibinfo{author}{\bibfnamefont{A.}~\bibnamefont{Burrows}},
  \bibinfo{journal}{Rev. Mod. Phys.} \textbf{\bibinfo{volume}{85}},
  \bibinfo{pages}{245} (\bibinfo{year}{2013}),
  \urlprefix\url{http://link.aps.org/doi/10.1103/RevModPhys.85.245}.

\bibitem[{\citenamefont{{Blondin} et~al.}(2003)\citenamefont{{Blondin},
  {Mezzacappa}, and {DeMarino}}}]{blondin:03}
\bibinfo{author}{\bibfnamefont{J.~M.} \bibnamefont{{Blondin}}},
  \bibinfo{author}{\bibfnamefont{A.}~\bibnamefont{{Mezzacappa}}},
  \bibnamefont{and}
  \bibinfo{author}{\bibfnamefont{C.}~\bibnamefont{{DeMarino}}},
  \bibinfo{journal}{\apj} \textbf{\bibinfo{volume}{584}}, \bibinfo{pages}{971}
  (\bibinfo{year}{2003}), \eprint{astro-ph/0210634}.

\bibitem[{\citenamefont{{Foglizzo} et~al.}(2007)\citenamefont{{Foglizzo},
  {Galletti}, {Scheck}, and {Janka}}}]{foglizzo:07}
\bibinfo{author}{\bibfnamefont{T.}~\bibnamefont{{Foglizzo}}},
  \bibinfo{author}{\bibfnamefont{P.}~\bibnamefont{{Galletti}}},
  \bibinfo{author}{\bibfnamefont{L.}~\bibnamefont{{Scheck}}}, \bibnamefont{and}
  \bibinfo{author}{\bibfnamefont{H.-T.} \bibnamefont{{Janka}}},
  \bibinfo{journal}{\apj} \textbf{\bibinfo{volume}{654}}, \bibinfo{pages}{1006}
  (\bibinfo{year}{2007}), \eprint{astro-ph/0606640}.

\bibitem[{\citenamefont{{Foglizzo} et~al.}(2006)\citenamefont{{Foglizzo},
  {Scheck}, and {Janka}}}]{foglizzo:06}
\bibinfo{author}{\bibfnamefont{T.}~\bibnamefont{{Foglizzo}}},
  \bibinfo{author}{\bibfnamefont{L.}~\bibnamefont{{Scheck}}}, \bibnamefont{and}
  \bibinfo{author}{\bibfnamefont{H.-T.} \bibnamefont{{Janka}}},
  \bibinfo{journal}{\apj} \textbf{\bibinfo{volume}{652}}, \bibinfo{pages}{1436}
  (\bibinfo{year}{2006}), \eprint{astro-ph/0507636}.

\bibitem[{\citenamefont{{Scheck} et~al.}(2008)\citenamefont{{Scheck}, {Janka},
  {Foglizzo}, and {Kifonidis}}}]{scheck:08}
\bibinfo{author}{\bibfnamefont{L.}~\bibnamefont{{Scheck}}},
  \bibinfo{author}{\bibfnamefont{H.-T.} \bibnamefont{{Janka}}},
  \bibinfo{author}{\bibfnamefont{T.}~\bibnamefont{{Foglizzo}}},
  \bibnamefont{and}
  \bibinfo{author}{\bibfnamefont{K.}~\bibnamefont{{Kifonidis}}},
  \bibinfo{journal}{\aap} \textbf{\bibinfo{volume}{477}}, \bibinfo{pages}{931}
  (\bibinfo{year}{2008}), \eprint{0704.3001}.

\bibitem[{\citenamefont{{Hempel} et~al.}(2012)\citenamefont{{Hempel},
  {Fischer}, {Schaffner-Bielich}, and
  {Liebend{\"o}rfer}}}]{2012ApJ...748...70H}
\bibinfo{author}{\bibfnamefont{M.}~\bibnamefont{{Hempel}}},
  \bibinfo{author}{\bibfnamefont{T.}~\bibnamefont{{Fischer}}},
  \bibinfo{author}{\bibfnamefont{J.}~\bibnamefont{{Schaffner-Bielich}}},
  \bibnamefont{and}
  \bibinfo{author}{\bibfnamefont{M.}~\bibnamefont{{Liebend{\"o}rfer}}},
  \bibinfo{journal}{\apj} \textbf{\bibinfo{volume}{748}}, \bibinfo{eid}{70}
  (\bibinfo{year}{2012}), \eprint{1108.0848}.

\bibitem[{\citenamefont{{Steiner} et~al.}(2013)\citenamefont{{Steiner},
  {Hempel}, and {Fischer}}}]{2013ApJ...774...17S}
\bibinfo{author}{\bibfnamefont{A.~W.} \bibnamefont{{Steiner}}},
  \bibinfo{author}{\bibfnamefont{M.}~\bibnamefont{{Hempel}}}, \bibnamefont{and}
  \bibinfo{author}{\bibfnamefont{T.}~\bibnamefont{{Fischer}}},
  \bibinfo{journal}{\apj} \textbf{\bibinfo{volume}{774}}, \bibinfo{eid}{17}
  (\bibinfo{year}{2013}), \eprint{1207.2184}.

\bibitem[{\citenamefont{{Smith} et~al.}(2011)\citenamefont{{Smith}, {Li},
  {Filippenko}, and {Chornock}}}]{2011MNRAS.412.1522S}
\bibinfo{author}{\bibfnamefont{N.}~\bibnamefont{{Smith}}},
  \bibinfo{author}{\bibfnamefont{W.}~\bibnamefont{{Li}}},
  \bibinfo{author}{\bibfnamefont{A.~V.} \bibnamefont{{Filippenko}}},
  \bibnamefont{and}
  \bibinfo{author}{\bibfnamefont{R.}~\bibnamefont{{Chornock}}},
  \bibinfo{journal}{\mnras} \textbf{\bibinfo{volume}{412}},
  \bibinfo{pages}{1522} (\bibinfo{year}{2011}), \eprint{1006.3899}.

\bibitem[{\citenamefont{{Ott} et~al.}(2006)\citenamefont{{Ott}, {Burrows},
  {Thompson}, {Livne}, and {Walder}}}]{ott:06spin}
\bibinfo{author}{\bibfnamefont{C.~D.} \bibnamefont{{Ott}}},
  \bibinfo{author}{\bibfnamefont{A.}~\bibnamefont{{Burrows}}},
  \bibinfo{author}{\bibfnamefont{T.~A.} \bibnamefont{{Thompson}}},
  \bibinfo{author}{\bibfnamefont{E.}~\bibnamefont{{Livne}}}, \bibnamefont{and}
  \bibinfo{author}{\bibfnamefont{R.}~\bibnamefont{{Walder}}},
  \bibinfo{journal}{\apjs} \textbf{\bibinfo{volume}{164}}, \bibinfo{pages}{130}
  (\bibinfo{year}{2006}).

\bibitem[{\citenamefont{{Lattimer} and {Douglas Swesty}}(1991)}]{lattimer:91}
\bibinfo{author}{\bibfnamefont{J.~M.} \bibnamefont{{Lattimer}}}
  \bibnamefont{and} \bibinfo{author}{\bibfnamefont{F.}~\bibnamefont{{Douglas
  Swesty}}}, \bibinfo{journal}{Nuclear Physics A}
  \textbf{\bibinfo{volume}{535}}, \bibinfo{pages}{331} (\bibinfo{year}{1991}).

\bibitem[{\citenamefont{{Veitch} et~al.}(2015)\citenamefont{{Veitch},
  {Raymond}, {Farr}, {Farr}, {Graff}, {Vitale}, {Aylott}, {Blackburn},
  {Christensen}, {Coughlin} et~al.}}]{veitch:15}
\bibinfo{author}{\bibfnamefont{J.}~\bibnamefont{{Veitch}}},
  \bibinfo{author}{\bibfnamefont{V.}~\bibnamefont{{Raymond}}},
  \bibinfo{author}{\bibfnamefont{B.}~\bibnamefont{{Farr}}},
  \bibinfo{author}{\bibfnamefont{W.}~\bibnamefont{{Farr}}},
  \bibinfo{author}{\bibfnamefont{P.}~\bibnamefont{{Graff}}},
  \bibinfo{author}{\bibfnamefont{S.}~\bibnamefont{{Vitale}}},
  \bibinfo{author}{\bibfnamefont{B.}~\bibnamefont{{Aylott}}},
  \bibinfo{author}{\bibfnamefont{K.}~\bibnamefont{{Blackburn}}},
  \bibinfo{author}{\bibfnamefont{N.}~\bibnamefont{{Christensen}}},
  \bibinfo{author}{\bibfnamefont{M.}~\bibnamefont{{Coughlin}}},
  \bibnamefont{et~al.}, \bibinfo{journal}{\prd} \textbf{\bibinfo{volume}{91}},
  \bibinfo{eid}{042003} (\bibinfo{year}{2015}), \eprint{1409.7215}.

\bibitem[{\citenamefont{{LIGO Algorithm Library}}()}]{LAL}
\bibinfo{author}{\bibnamefont{{LIGO Algorithm Library}}},
  \emph{\bibinfo{title}{www.lsc-group.phys.uwm.edu/daswg/projects/lalsuite.html}},
  \urlprefix\url{www.lsc-group.phys.uwm.edu/daswg/projects/lalsuite.html}.

\bibitem[{\citenamefont{{IceCube Collaboration}
  et~al.}(2006)\citenamefont{{IceCube Collaboration}, {Achterberg},
  {Ackermann}, {Adams}, {Ahrens}, {Andeen}, {Atlee}, {Baccus}, {Bahcall}, {Bai}
  et~al.}}]{2006APh....26..155I}
\bibinfo{author}{\bibnamefont{{IceCube Collaboration}}},
  \bibinfo{author}{\bibfnamefont{A.}~\bibnamefont{{Achterberg}}},
  \bibinfo{author}{\bibfnamefont{M.}~\bibnamefont{{Ackermann}}},
  \bibinfo{author}{\bibfnamefont{J.}~\bibnamefont{{Adams}}},
  \bibinfo{author}{\bibfnamefont{J.}~\bibnamefont{{Ahrens}}},
  \bibinfo{author}{\bibfnamefont{K.}~\bibnamefont{{Andeen}}},
  \bibinfo{author}{\bibfnamefont{D.~W.} \bibnamefont{{Atlee}}},
  \bibinfo{author}{\bibfnamefont{J.}~\bibnamefont{{Baccus}}},
  \bibinfo{author}{\bibfnamefont{J.~N.} \bibnamefont{{Bahcall}}},
  \bibinfo{author}{\bibfnamefont{X.}~\bibnamefont{{Bai}}},
  \bibnamefont{et~al.}, \bibinfo{journal}{Astroparticle Physics}
  \textbf{\bibinfo{volume}{26}}, \bibinfo{pages}{155} (\bibinfo{year}{2006}),
  \eprint{astro-ph/0604450}.

\bibitem[{\citenamefont{{Ageron} et~al.}(2011)\citenamefont{{Ageron},
  {Aguilar}, {Al Samarai}, {Albert}, {Ameli}, {Andr{\'e}}, {Anghinolfi},
  {Anton}, {Anvar}, {Ardid} et~al.}}]{2011NIMPA.656...11A}
\bibinfo{author}{\bibfnamefont{M.}~\bibnamefont{{Ageron}}},
  \bibinfo{author}{\bibfnamefont{J.~A.} \bibnamefont{{Aguilar}}},
  \bibinfo{author}{\bibfnamefont{I.}~\bibnamefont{{Al Samarai}}},
  \bibinfo{author}{\bibfnamefont{A.}~\bibnamefont{{Albert}}},
  \bibinfo{author}{\bibfnamefont{F.}~\bibnamefont{{Ameli}}},
  \bibinfo{author}{\bibfnamefont{M.}~\bibnamefont{{Andr{\'e}}}},
  \bibinfo{author}{\bibfnamefont{M.}~\bibnamefont{{Anghinolfi}}},
  \bibinfo{author}{\bibfnamefont{G.}~\bibnamefont{{Anton}}},
  \bibinfo{author}{\bibfnamefont{S.}~\bibnamefont{{Anvar}}},
  \bibinfo{author}{\bibfnamefont{M.}~\bibnamefont{{Ardid}}},
  \bibnamefont{et~al.}, \bibinfo{journal}{Nuclear Instruments and Methods in
  Physics Research A} \textbf{\bibinfo{volume}{656}}, \bibinfo{pages}{11}
  (\bibinfo{year}{2011}), \eprint{1104.1607}.

\bibitem[{\citenamefont{{Klimenko} et~al.}(2016)}]{2016PhRvD..93d2004K}
\bibinfo{author}{\bibfnamefont{S.}~\bibnamefont{{Klimenko}}}
  \bibnamefont{et~al.}, \bibinfo{journal}{\prd} \textbf{\bibinfo{volume}{93}},
  \bibinfo{eid}{042004} (\bibinfo{year}{2016}), \eprint{1511.05999}.

\bibitem[{\citenamefont{Klimenko et~al.}(2008)}]{0264-9381-25-11-114029}
\bibinfo{author}{\bibfnamefont{S.}~\bibnamefont{Klimenko}}
  \bibnamefont{et~al.}, \bibinfo{journal}{Classical and Quantum Gravity}
  \textbf{\bibinfo{volume}{25}}, \bibinfo{pages}{114029}
  (\bibinfo{year}{2008}),
  \urlprefix\url{http://stacks.iop.org/0264-9381/25/i=11/a=114029}.

\bibitem[{\citenamefont{{Lynch} et~al.}(2015)\citenamefont{{Lynch}, {Vitale},
  {Essick}, {Katsavounidis}, and {Robinet}}}]{lynch:15}
\bibinfo{author}{\bibfnamefont{R.}~\bibnamefont{{Lynch}}},
  \bibinfo{author}{\bibfnamefont{S.}~\bibnamefont{{Vitale}}},
  \bibinfo{author}{\bibfnamefont{R.}~\bibnamefont{{Essick}}},
  \bibinfo{author}{\bibfnamefont{E.}~\bibnamefont{{Katsavounidis}}},
  \bibnamefont{and}
  \bibinfo{author}{\bibfnamefont{F.}~\bibnamefont{{Robinet}}},
  \bibinfo{journal}{arXiv:1511.05955}  (\bibinfo{year}{2015}),
  \eprint{1511.05955}.

\bibitem[{\citenamefont{Necula et~al.}(2012)\citenamefont{Necula, Klimenko, and
  Mitselmakher}}]{1742-6596-363-1-012032}
\bibinfo{author}{\bibfnamefont{V.}~\bibnamefont{Necula}},
  \bibinfo{author}{\bibfnamefont{S.}~\bibnamefont{Klimenko}}, \bibnamefont{and}
  \bibinfo{author}{\bibfnamefont{G.}~\bibnamefont{Mitselmakher}},
  \bibinfo{journal}{Journal of Physics: Conference Series}
  \textbf{\bibinfo{volume}{363}}, \bibinfo{pages}{012032}
  (\bibinfo{year}{2012}),
  \urlprefix\url{http://stacks.iop.org/1742-6596/363/i=1/a=012032}.

\bibitem[{\citenamefont{{Klimenko} et~al.}(2005)\citenamefont{{Klimenko},
  {Mohanty}, {Rakhmanov}, and {Mitselmakher}}}]{2005PhRvD..72l2002K}
\bibinfo{author}{\bibfnamefont{S.}~\bibnamefont{{Klimenko}}},
  \bibinfo{author}{\bibfnamefont{S.}~\bibnamefont{{Mohanty}}},
  \bibinfo{author}{\bibfnamefont{M.}~\bibnamefont{{Rakhmanov}}},
  \bibnamefont{and}
  \bibinfo{author}{\bibfnamefont{G.}~\bibnamefont{{Mitselmakher}}},
  \bibinfo{journal}{\prd} \textbf{\bibinfo{volume}{72}}, \bibinfo{eid}{122002}
  (\bibinfo{year}{2005}), \eprint{gr-qc/0508068}.

\bibitem[{\citenamefont{{Veitch} and {Vecchio}}(2010)}]{2010PhRvD..81f2003V}
\bibinfo{author}{\bibfnamefont{J.}~\bibnamefont{{Veitch}}} \bibnamefont{and}
  \bibinfo{author}{\bibfnamefont{A.}~\bibnamefont{{Vecchio}}},
  \bibinfo{journal}{\prd} \textbf{\bibinfo{volume}{81}}, \bibinfo{eid}{062003}
  (\bibinfo{year}{2010}), \eprint{0911.3820}.

\bibitem[{\citenamefont{{Abbott} et~al.}(2009)\citenamefont{{Abbott}, {Abbott},
  {Adhikari}, {Ajith}, {Allen}, {Allen}, {Amin}, {Anderson}, {Anderson},
  {Arain} et~al.}}]{2009RPPh...72g6901A}
\bibinfo{author}{\bibfnamefont{B.~P.} \bibnamefont{{Abbott}}},
  \bibinfo{author}{\bibfnamefont{R.}~\bibnamefont{{Abbott}}},
  \bibinfo{author}{\bibfnamefont{R.}~\bibnamefont{{Adhikari}}},
  \bibinfo{author}{\bibfnamefont{P.}~\bibnamefont{{Ajith}}},
  \bibinfo{author}{\bibfnamefont{B.}~\bibnamefont{{Allen}}},
  \bibinfo{author}{\bibfnamefont{G.}~\bibnamefont{{Allen}}},
  \bibinfo{author}{\bibfnamefont{R.~S.} \bibnamefont{{Amin}}},
  \bibinfo{author}{\bibfnamefont{S.~B.} \bibnamefont{{Anderson}}},
  \bibinfo{author}{\bibfnamefont{W.~G.} \bibnamefont{{Anderson}}},
  \bibinfo{author}{\bibfnamefont{M.~A.} \bibnamefont{{Arain}}},
  \bibnamefont{et~al.}, \bibinfo{journal}{Reports on Progress in Physics}
  \textbf{\bibinfo{volume}{72}}, \bibinfo{eid}{076901} (\bibinfo{year}{2009}),
  \eprint{0711.3041}.

\bibitem[{\citenamefont{{Vallisneri} et~al.}(2015)}]{vallisneri:15}
\bibinfo{author}{\bibfnamefont{M.}~\bibnamefont{{Vallisneri}}}
  \bibnamefont{et~al.}, \bibinfo{journal}{Journal of Physics Conference Series}
  \textbf{\bibinfo{volume}{610}}, \bibinfo{eid}{012021} (\bibinfo{year}{2015}),
  \eprint{1410.4839}.

\end{thebibliography}

\end{document}